\documentclass[11pt,twoside,a4paper,pdf]{article}
\usepackage{graphicx,color}
\usepackage{times}
\usepackage{booktabs}
\usepackage{cite}
\usepackage{a4wide}
\usepackage{xspace}
\usepackage{amsmath, amssymb}
\usepackage[small,bf]{caption} % from Dima
\usepackage[colorlinks=true, pdftex,  urlcolor=blue, 
  citecolor=blue, pdftitle={Helicity-Dependent
    Showers and Matching with VINCIA}, pdfauthor={A. Larkoski,
    J.J. Lopez-Villarejo and Peter Zeiler Skands}, pdfsubject={Monte
    Carlo generators}, pdfkeywords={Monte Carlo, Event Generators, QCD,
    Parton Showers, Matching, Jets, VINCIA}]{hyperref}
\usepackage{subfig}
\usepackage{slashed}

\newcommand{\eqRef}[1]{eq.~\eqref{#1}\xspace}

\newcommand{\appRef}[1]{appendix~\ref{#1}\xspace}

\newcommand{\secRef}[1]{section~\ref{#1}\xspace}

\newcommand{\tabRef}[1]{tab.~\ref{#1}\xspace}

\newcommand{\figRef}[1]{fig.~\ref{#1}\xspace}
\newcommand{\FigRef}[1]{Fig.~\ref{#1}\xspace}

\newcommand{\mrm}[1]{\mathrm{#1}}

\newcommand{\ttt}[1]{\texttt{#1}}

\newcommand{\TeV}{\,\mbox{Te\kern-0.2exV}}
\newcommand{\GeV}{\,\mbox{Ge\kern-0.2exV}}
\newcommand{\MeV}{\,\mbox{Me\kern-0.2exV}}
\newcommand{\keV}{\,\mbox{ke\kern-0.2exV}}
\newcommand{\eV}{\,\mbox{e\kern-0.2exV}}

\newcommand{\MC}[1]{{#1}}

\newcommand{\Ar}{\MC{ARIADNE}\xspace}

\newcommand{\Mg}{\MC{MADGRAPH}\xspace}
\newcommand{\Py}{\MC{PYTHIA}\xspace}
\newcommand{\Sh}{\MC{SHERPA}\xspace}
\newcommand{\Vc}{\MC{VINCIA}\xspace}

\begin{document}

\begin{minipage}{\textwidth}\vspace*{-5mm}
\flushright
CERN-PH-TH/2012-257\\
MIT-CTP 4406\\
FTUAM-12-105\\
IFT-UAM/CSIC-12-96\\

\end{minipage}
\vskip8mm
\begin{center}
\Large{\bf Helicity-Dependent Showers and Matching with VINCIA}\\[5mm]
\end{center}
\vskip5mm
\begin{center}
{\large A.~J.~Larkoski$^1$, J.J.~Lopez-Villarejo$^{2,3}$, and
  P.~Skands$^2$}\\[2mm]\small 
\begin{minipage}{0.8\textwidth}
$^1$: Center for Theoretical Physics, Massachusetts Institute of Technology, Cambridge, MA 02139 U.S.A. \\
$^2$: Theoretical Physics, CERN CH-1211, Geneva 23, Switzerland\\
$^3$: Univ. Autonoma de Madrid and IFT-UAM/CSIC, Madrid 28049, Spain
\end{minipage}
\end{center}
%\maketitle
\vskip5mm
\begin{center}
\parbox{0.83\textwidth}{\small
\textbf{Abstract} --- 
We present an antenna-shower formalism that includes helicity
dependence for massless partons. The formalism applies to both
traditional (global) showers and to sector-based
variants. We combine the shower with VINCIA's multiplicative approach to 
matrix-element matching, generalized to operate on each helicity 
configuration separately. The result is a substantial gain in
computational speed for high parton multiplicities. We present an 
implementation of both sector and global showers, with min/max variations, and
helicity-dependent tree-level matching applied through $n\le 4$ for
$V/H\to q\bar{q}+n\,\mbox{partons}$ and through $n\le 5$ for 
$H\to n\,\mrm{gluons}$. 
\vspace*{3mm}
}
\end{center}
\vskip8mm

% INTRODUCTION
\section{Introduction}
Multi-leg amplitudes and their combination with parton-shower
resummations, called ME/PS matching, are among the most active topics in
current high-energy phenomenology (see \cite{Buckley:2011ms} for a review). 
What ME/PS matching provides is a calculation that smoothly 
interpolates between fixed-order QCD (and QED) amplitudes in the 
high-$p_\perp$ region and infinite-order logarithmic approximations
in the low-$p_\perp$ one. Importantly, the output of such
calculations is in the form of fully hadronized ``events'', which 
can be subjected to direct and detailed experimental
comparison. 

However, current state-of-the-art 
multileg ME/PS methods, such as
CKKW-L~\cite{Catani:2001cc,Lonnblad:2001iq}, 
MLM~\cite{Mangano:2006rw}, and MENLOPS~\cite{Hamilton:2010wh}, 
are rather computationally intensive,
so that the most complex calculations can only be carried out on large
clusters. The increase in computational time with the number of legs
is partly due to the amplitudes becoming more complicated at each
order, but more importantly these methods (which we refer to 
collectively as ``slicing'' methods~\cite{Skands:2012ts})
algorithmically treat each multi-leg matching
step as completely unrelated to all the others: a separate phase-space
integration, event-generation, and event unweighting step is needed for
each multiplicity. As shown in \cite{LopezVillarejo:2011ap}, much
faster algorithms can be constructed by ``nesting'' the successive
matching steps within each other, starting from the Born level and
using an overestimating (``trial'') parton shower as the only
additional phase-space generator. The matrix-element amplitudes can then be
imprinted on the final answer by a simple Monte-Carlo
veto step. This 
approach, which we refer to as the ``multiplicative'' method, 
was first developed for one additional leg in~\cite{Bengtsson:1986hr} 
and was generalized to multiple legs in
\cite{Giele:2011cb}.

In this paper, we develop an additional refinement of the
multiplicative method, which further increases the algorithmic speed
that can be 
achieved when matching to large parton multiplicities. As a beneficial
side effect, the intrinsic precision of the underlying parton-shower
formalism is increased as well. The main point is to replace the
ordinary (helicity-summed) shower radiation functions with
helicity-dependent ones, such as those given 
in~\cite{Larkoski:2009ah}. That is, we shall treat 
(massless) quarks and gluons with negative and positive helicities as
effectively being different particles. 
The resulting shower generates a LL 
approximation to each individual multi-leg helicity amplitude
(squared) separately, 
and the resulting evolution can therefore be matched to
one single such amplitude at a time. 
Thus, instead of taking sums and averages at each order, we are now
effectively sampling helicity space 
by Monte Carlo, saving substantial time when matching to several
successive legs. 

Note that, up to the matched orders, 
interference effects between amplitudes with
different internal helicity structures are still fully taken into account,
since the last matched helicity amplitude does contain a sum over all
contributing internal-line helicities. 
At subsequent orders, however, the helicity-dependent shower does not
generate the equivalent of a full 
spin-density matrix treatment (see e.g.~\cite{Richardson:2001df}). Nor
are any explicit azimuth-dependent correlations 
among successive emissions manifest in the helicity basis 
as would be the case with linearly polarized Altarelli-Parisi kernels \cite{stirling96}. The final 
precision is nevertheless still improved, since unphysical
helicity assignments are not allowed to contribute.

In \secRef{sec:shower}, we generalize the \Vc shower and matching
formalism~\cite{Giele:2007di} to include helicity dependence.
In \secRef{sec:formalism}, we derive explicit helicity-dependent QCD antenna
functions, considering both sector and global antenna types. 
 Finally, in \secRef{sec:results} we present a set of comparisons to
matrix elements, speed benchmarks, and validations on selected LEP
distributions, and in \secRef{sec:conclusion} we summarize and
provide a brief outlook.

% FORMALISM

\section{The Shower and Matching Algorithm \label{sec:shower}}

The helicity of a particle is the projection of the spin of the particle
onto its momentum.  For massless particles, 
helicity is Lorentz invariant 
and takes the values $\pm s$ for particles with total spin $s$.  Typically,
in computing matrix elements, one sums over helicities of the incoming
and outgoing particles because they are not directly observed.  
However, beginning with observations by Parke and Taylor \cite{Parke:1986gb} in the 1980s,
it was discovered that individual helicity amplitudes are significantly
simpler in form than are helicity-summed matrix elements.  In addition, chirality
or handedness is important in processes mediated by the weak interaction.
Thus, we expect that a Monte Carlo parton shower based on the helicity
structure of QCD, rather than summed over helicities, is both faster when matching to matrix
elements as well as more accurate, especially in weak decays.  

In this section, we discuss the modifications to the \Vc shower and matching 
algorithms required to take helicity into account. The antenna
functions themselves will be the topic of \secRef{sec:formalism}.

\subsection{Helicity-Dependent Showering}

The helicity-dependent shower algorithm follows the
unpolarized one quite closely~\cite{Giele:2007di}, with differences
entering only in a few very specific places, as follows. 

If \Vc is asked to shower an event that contains unpolarized partons,
e.g.\ an unpolarized $Z\to q\bar{q}$ event, it first hands
the event to a 
\emph{polarizer} function which searches the 
\Vc library for helicity-dependent matrix elements
corresponding to the given process (see \secRef{sec:matching} below
for a full list). If this search is successful, helicities
for the final-state partons are assigned based on their relative
matrix-element weights.
In the example of on-shell $Z\to d\bar{d}$ ($Z\to
u\bar{u}$) decay, a phase-space-dependent average of 97\% (83\%) of the events will end up as $q_L\bar{q}_R$,
with the other being $q_R\bar{q}_L$.  For off-shell
$Z$ bosons, the full $e^+ e^-\to \gamma^*/Z \to q\bar{q}$ matrix
elements would be used instead. 

Trial branchings are generated as in the unpolarized shower, 
according to the unpolarized trial
functions.  The unpolarized trial functions are essentially an overestimating eikonal term, see
\cite{Giele:2011cb}, with additional collinear-singular terms for
sector showers~\cite{LopezVillarejo:2011ap}. 
After the selection of branching
invariants, the probability that the branching will be accepted at
all, summed over all possible post-branching helicities, is computed:
\begin{equation}
P_{\mrm{accept}} \ = \ \frac{a_\mrm{phys}}{a_\mrm{trial}} \ =
\ \frac{\sum_{h_i,h_j,h_k} a(h_A,h_B\to h_i, h_j, h_k)}{a_\mrm{trial}}~,
\label{eq:accept}
\end{equation}
with $h_{A,B}$ the (fixed) helicities of the parent partons,
$h_{i,j,k}$ the helicities of the daughter ones, and $a(h_A,h_B\to
h_i, h_j, h_k)$ a helicity-dependent antenna function, the precise
forms of which will be discussed in \secRef{sec:formalism}. This accept
probability is not exactly identical to the unpolarized one, since
the helicities of the parent partons $A$ and $B$ are not averaged
over. Note also that, for sector showers, the trial function appearing
in the denominator should be the full one containing both the
soft-eikonal and the additional collinear-singular trial terms. 

Helicities for the three
daughter partons are then assigned according to the relative
probabilities
\begin{equation}
P(h_A,h_B\to h_i,h_j,h_k) \ = \ \frac{a(h_A,h_B\to
  h_i,h_j,h_k)}{\sum_{h_i,h_j,h_k} a(h_A,h_B\to h_i, h_j, h_k)}~.
\end{equation}
Note that it is important that the denominator here be exactly the
same as the numerator in \eqRef{eq:accept}.

All other aspects of the showering remain unmodified (for matching,
see below). For
completeness, we also note that, when helicity dependence is switched
off, the unpolarized antenna functions are obtained as direct helicity
sums over the helicity-dependent ones, averaging over the parent
helicities. The treatments with and without helicity dependence are
thus intimately related and it is straightforward to go from one to
the other. 

For one showering step, there should therefore also be little difference
between the helicity-dependent and unpolarized treatments, up to
differences caused by the helicity-dependent finite terms not being
equal to their averages. However, the chosen helicities at one step then
become the input helicities 
  for the next step. This excludes some unphysical configurations from
  the effective helicity average/sum in the next step, yielding an
  improvement in accuracy over the unpolarized case.

\subsection{Matching to Matrix Elements \label{sec:matching}}

The procedure for matching helicity matrix elements to the helicity-dependent
parton shower in \Vc is similar for matching to spin-summed matrix elements~\cite{Giele:2011cb}.
At each step in the shower, the parton shower provides an estimate for
the ratio 
\begin{equation}
P_n=\frac{|{\cal M}_{n}|^2}{|{\cal M}_{n-1}|^2} \ ,
\end{equation}
where $|{\cal M}_{n}|^2$ is the matrix element after $n$ additional
quark or gluon emissions 
from the Born-level matrix element.  
To match the parton shower to the exact matrix element at this stage, the emission of the $n$th parton is accepted
with the probability
\begin{equation}\label{eq:matchaccept}
P^n_{\mrm{accept}} = \frac{a^{\mrm{Phys}}}{a^{\mrm{Trial}}} P_{\mrm{ME}} \ .
\end{equation}
$a^{\mrm{Phys}}$ is the antenna for the current branching and $a^{\mrm{Trial}}$ is an overestimate
antenna that is strictly larger than $a^{\mrm{Phys}}$.  The
matching factor $P_{\mrm{ME}}$ is
\begin{equation}
P_{\mrm{ME}} = \frac{|{\cal M}_n|^2}{ |{\cal M}_n|^2_{\mrm{shower}}} , 
\end{equation}
where $|{\cal M}_{n}|^2_{\mrm{shower}}$ is the approximation to the
matrix element provided by the parton shower:
\begin{equation}
|{\cal M}_n|^2_{\mrm{shower}} = \sum_{i} a_i |{\cal M}_{n-1 \ i}|^2 \ ,
\end{equation}
with $i$ running over the possible clusterings and helicities and
$a_i$ the corresponding
antenna for the $i$th configuration.
Note that the form of this approximation depends on the formulation of the
shower (global vs.~sector); see~\cite{LopezVillarejo:2011ap} for
details on the differences between global and sector matching.

The ratio $P_n$ in a helicity-dependent shower has especially nice and general
properties. For decays of colorless resonances to massless quarks, %or gluons, 
$P_n$ is independent of the $CP$ structure of the resonance for
helicity matrix elements. This allows the use of, for example, matrix
elements for $Z$ decay to massless quarks 
for use in matching processes which include $W$ bosons or photons.
Only the total spin of the 
resonance is relevant for matching to helicity matrix elements.

The proof of this is straightforward.  Consider the ratio of amplitudes
\begin{equation}
\rho\equiv\frac{{\cal M}_{n}}{{\cal M}_{n-1}} \ .
\end{equation}
This can be written generically as
\begin{equation}
\rho= \frac{{\cal M}_{n}}{{\cal M}_{n-1}}=\frac{\bar{u}(q_f){\cal O}_g^n 
\slashed{J}(a-b\gamma_5){\cal O}_g^nu(\bar{q}_f)}{\bar{u}(q_i){\cal O}_g^{n-1} 
\slashed{J}(a-b\gamma_5){\cal O}_g^{n-1}u(\bar{q}_i)} \ ,
\end{equation}
where $q_i,q_f$ ($\bar{q}_i,\bar{q}_f$) are the (anti)quark before and after additional 
radiation, respectively, $\slashed{J}$ is the current carried by the resonance, 
$(a-b\gamma_5)$ parametrizes the $CP$ structure of the current, and ${\cal O}_g^n$ 
is an unspecified operator which produces additional QCD radiation up to ${\cal O}(g_s^n)$.  Irrespective of 
the form of the current $\slashed{J}$, the radiation operators ${\cal O}_g^n$ and ${\cal O}_g^{n-1}$ contain an 
even number of gamma matrices: one for each quark-gluon vertex and one for each 
quark propagator.  Then, it is true that 
\begin{equation}
[{\cal O}_g^n,\gamma_5]=[{\cal O}_g^{n-1},\gamma_5]=0 \ ,
\end{equation}
which allows us to freely interchange the position of $(a-b\gamma_5)$ and ${\cal O}_g$.  
We then have
\begin{equation}
\rho= \frac{{\cal M}_{n}}{{\cal M}_{n-1}}=\frac{\bar{u}(q_f){\cal O}_g^n 
\slashed{J}{\cal O}_g^n(a-b\gamma_5)u(\bar{q}_f)}{\bar{u}(q_i) {\cal O}_g^{n-1}\slashed{J}{\cal O}_g^{n-1}(a-b\gamma_5)u(\bar{q}_i)} \ .
\end{equation}
Helicity spinors are eigenvectors of $\gamma_5$ with eigenvalue equal to their 
helicity.  The helicity of massless quarks is preserved in QCD; that is,
\begin{equation}
\gamma_5u(\bar{q})=h_{\bar{q}}u(\bar{q}) \qquad \text{and} \qquad h_{\bar{q}_i} = h_{\bar{q}_f} \ .
\end{equation}
We then find that 
\begin{equation}
\rho=\frac{\bar{u}(q_f){\cal O}_g^n \slashed{J}{\cal O}_g^nu(\bar{q}_f)(a-bh_{\bar{q}})}{\bar{u}(q_i) {\cal O}_g^{n-1}
\slashed{J}{\cal O}_g^{n-1}u(\bar{q}_i)(a-bh_{\bar{q}})} =\frac{\bar{u}(q_f){\cal O}_g^n \slashed{J}{\cal O}_g^nu(\bar{q}_f)}{\bar{u}(q_i){\cal O}_g^{n-1}
 \slashed{J}{\cal O}_g^{n-1}u(\bar{q}_i)} \ ,
\end{equation}
proving that $P_n$ is independent of the $CP$ structure of the resonance.
Note that by crossing symmetry, this proof also holds for particles created in the final state from parton-parton
scattering.

Based on the universality of $P_n$,  
the current \Vc implementation includes matching to the helicity
amplitudes listed in \tabRef{tab:isMatched}, with $V$ and $S$ denoting
a generic colorless spin-1 and spin-0 particle, respectively. 
\begin{table}[t]
\centering
\begin{tabular}{c|c|c|c|c|c}
Decaying & \multicolumn{4}{c}{Order Beyond Born}\\
Particle & 0 & 1 & 2 & 3 & 4 \\\toprule
$V,S$ & $q\bar{q}$ & $q\bar{q}g$ & $q\bar{q}gg$, $q\bar{q}q'\bar{q}'$ & $q\bar{q}ggg$, $q\bar{q}q\bar{q}'g$
& $q\bar{q}gggg$, $q\bar{q}q'\bar{q}'gg$, $q\bar{q}q'\bar{q}'q''\bar{q}''$ \\
%$S\to q\bar{q}$ & +$g$ & +$gg$, +$q'\bar{q}'$ & +$ggg$, +$gq\bar{q}'$
%& +$gggg$ \\
$S$ & $gg$ & $ggg$ & $gggg$ & $ggggg$ & - \\\bottomrule
\end{tabular}
\caption{Matrix elements available in \Vc for helicity-dependent
  matching corrections, with $V$ ($S$) a generic colorless spin-1
  (spin-0) boson with arbitrary couplings.\label{tab:isMatched}}
\end{table}
The corresponding massless helicity amplitudes squared were obtained by modifying \Mg
v.~4.4.26~\cite{Alwall:2007st} to extract individual helicity configurations and are
evaluated at runtime using the HELAS libraries~\cite{Murayama:1992gi}.  
For processes involving decaying massive vector bosons, the spin of the vector is summed over
in computing the helicity matrix element.  In addition, matching is done for on-shell
particles and so the mass of the decaying resonance is artificially set to the 
center of mass energy $Q$.
Note that the HELAS libraries are so far not included in the \Vc package 
itself and must be downloaded separately. The default 
\Vc \ttt{make} target will automatically attempt to download the \Mg
package and compile the HELAS library from there. Alternatively, the user 
has the option of providing a precompiled HELAS library. 

\subsection{Massive Partons}\label{sec:mass}
The helicity-dependent formalism presented in this paper is limited to
massless partons. For massive partons, \Vc reverts
to the unpolarized massive framework presented in
\cite{GehrmannDeRidder:2011dm}. By default, this applies to charm and
heavier quarks. User options are provided that allow the massive treatment
to be applied also to $s$ quarks, or to force  
$c$ and/or $b$ ones to be treated as massless.
For completeness, we describe here what
the code does when helicity-dependence is switched on (as it is by
default) and one or more massive partons are present either in the
Born-level event (e.g., via a $Z\to b\bar{b}$ decay) or are created
during the shower evolution (e.g., via a $g\to b\bar{b}$ splitting).

 If a massive parton is present already in the Born-level event,
  the entire event is treated as unpolarized. That is, all parton
  helicities are ignored, including those of massless partons. The
  massive shower algorithm described in \cite{GehrmannDeRidder:2011dm}
  is applied, and hence mass corrections are included, even if
  helicity-dependence is not. 

If no massive parton is present in the Born-level event, the
helicity-dependent shower described in this paper is applied to the
event, but trial splittings of gluons to massive quarks are still
allowed. If such a branching is accepted, all helicities are then
ignored \emph{from that point onwards}, and the further event
evolution proceeds according to the unpolarized
algorithm~\cite{GehrmannDeRidder:2011dm}, as above. 

A subtlety arises concerning the matching to matrix elements. As
described above, the helicity-dependent formalism allows us to use
matrix elements for $Z$ decay to represent any vector boson, and ones
for $H$ decay to represent any scalar. For spin-summed matrix
elements, however, this universality breaks down. The user should
therefore be aware that, while
the full range of matched matrix elements are still
available for $Z$ and $H^0$ decays to unpolarized massive particles,
the corresponding corrections for $W$ and $H^+$ decays to massive
partons have so far not been implemented. 

For future reference, we 
note that phase-space maps for antennae involving massive particles
are available in \cite{GehrmannDeRidder:2011dm} and a set of 
spin-dependent antenna functions were defined in
\cite{Larkoski:2011fd}.

\section{Helicity-Dependent Antenna Functions \label{sec:formalism}}

Parton showers (including the dipole/antenna varieties) 
are governed by the properties of
soft and collinear emissions 
in QCD.  
The soft and collinear limits of massless QCD matrix elements are the universal
Altarelli-Parisi splitting functions \cite{Altarelli:1977zs} and take the schematic form
\begin{equation}
\lim_{s_{ij}\to 0}|{\cal M}(1,\dotsc,i,j,\dotsc,n)|^2=\frac{1}{s_{ij}}g_s^2{\cal C}_{ij} P_{i ,j \leftarrow \widehat{ij}}(z)|{\cal M}(1,\dotsc,\widehat{ij},\dotsc,n)|^2 \ ,
\end{equation}
where ${\cal C}_{ij}$ is the color factor, $g_s^2=4\pi\alpha_s$ is the QCD coupling and particles $i$ and $j$ are replaced by 
$\widehat{ij}$ in the matrix element on the right.  $P_{i ,j \leftarrow \widehat{ij}}(z)$ is 
the splitting function representing the distribution of energy fraction $z$ carried
by particle $i$.  The Altarelli-Parisi splitting functions for massless quarks and gluons of
definite helicity were given in their original paper and are reproduced 
in~\tabRef{tab:Altarelli}.  Note that rows in~\tabRef{tab:Altarelli} sum to the
familiar, unpolarized, Altarelli-Parisi splitting functions.  

\begin{table}
\begin{center}
\begin{tabular}{c | cccc}
 & $++$ &  $-+$ & $+-$ & $--$ \\ \hline
$ g_+ \to gg$\ : 
 & $ 1 / z(1-z)$   &   $(1-z)^3/ z$ & $z^3 / (1-z)$ & 0 \\ 
$ g_+ \to q \bar q$\ :  
 &  -   &   $ (1-z)^2$ & $ z^2 $ & - \\ 
$ q_+ \to qg$\ :   
&    $1/(1-z)$    & - &  $z^2/(1-z)$ & - \\ 
$ q_+ \to gq$\ : 
 &   $1/z$   &  $(1-z)^2/z$ & -  & -\\ 
\end{tabular}
\caption{Helicity-dependent Altarelli-Parisi splitting functions $P(z)$
for splittings $a \to bc$, with $z$ defined as the 
energy fraction taken by parton $b$.  The 
labels in the top row denote the helicities of the two final
particles in the order 
they appear: $(h_b,h_c)$.  The empty columns are forbidden 
by quark chiral 
symmetry. By the P and C invariance of QCD, 
the same expressions
apply after exchanging $- \leftrightarrow +$ or $q\leftrightarrow \bar q$.
\label{tab:Altarelli}}
\end{center}
\end{table}

The \Vc Monte Carlo is a dipole-antenna shower \cite{Giele:2007di} based on nested
$2\to 3$ splitting processes.  This splitting can be represented as $IK\to ijk$,
for initial partons $I$, $K$ and final partons $i$, $j$, $k$.  As \Vc works in the
color-ordered limit of QCD, the initial and final partons are assumed to be in
color order, as well.  We will also assume that all partons are massless, unless
otherwise specified.  The phase space for emission is defined by the dimensionless 
variables $y_{ij}$ and $y_{jk}$ where
\begin{equation}
y_{ij} = \frac{2 p_i \cdot p_j}{s} \ , \qquad y_{jk} = \frac{2 p_j \cdot p_k}{s} \ ,
\end{equation}
and $s\equiv (p_i+p_j+p_k)^2=(p_I+p_K)^2$ is the invariant mass of the dipole antenna
system.  The phase space of the emission is defined by the triangle
$y_{ij},y_{jk}\geq0$, $y_{ij}+y_{jk}\leq 1$.

The probability of emission is governed by the antenna function which is a function
of all relevant momenta, quantum numbers and the formulation of the shower.  For
the splitting $IK\to ijk$, the antenna function can be expressed in the form 
\begin{equation}
a^{\text{type(order)}}_{j/IK}(p_i,p_j,p_k) \ ,
\end{equation}
where type refers to global or sector antennae and order is the order in $\alpha_s$ to which the antennae
are computed.  When obvious from context, the superscripts will be omitted.
  In this paper, we will consider exclusively the lowest order antenna
functions and so we can define the color- and coupling- stripped antenna
\begin{equation}
a^{}_{j/IK}(p_i,p_j,p_k) = g_s^2 {\cal C}_{j/IK} \bar{a}_{j/IK}(p_i,p_j,p_k) \ .
\end{equation}
For simplicity, we will work with the color- and coupling-stripped antenna in the following.
For massless partons, $\bar{a}_{j/IK}(p_i,p_j,p_k)$ is a function of the kinematic invariants
$y_{ij}$ and $y_{jk}$ only.

The unpolarized global and sector antennae used in \Vc were defined in \cite{Giele:2007di,Giele:2011cb,LopezVillarejo:2011ap}.
We wish to extend the global and sector antennae to include full helicity dependence
of all partons in the antenna.  Our discussion will only include antennae in which all particles
are massless.  Antenna splitting functions including helicity dependence
were defined in \cite{Larkoski:2009ah} as ratios of matrix elements, but here, we will present a general treatment of the form of the
antennae.  There are many constraints that must be imposed on the antennae to determine
the singular terms; most importantly, the helicity-dependent antenna functions
must appropriately reproduce the helicity-dependent Altarelli-Parisi splitting functions
in the collinear limits.  Note that this only constrains the singular terms of the antenna;
the non-singular terms are unconstrained and can be interpreted as uncertainties in higher log-order terms.
Also, when summed over final parton helicities, the antenna functions should reproduce the
unpolarized antennae functions, up to terms that are non-singular.  
In the following subsections, we will discuss the construction of global and
sector helicity-dependent antennae.  

Before discussing the global and sector antennae, we will distinguish 
the definition and utility of helicity to define a massless particle's spin
from other definitions in the literature or used in simulation code.  The Les Houches
Accords of 2001~\cite{Boos:2001cv} outlined a set of variables by which to define the 
properties of particles in Monte Carlo event simulations.  The variable {\tt SPINUP}
was introduced to quantify the spin of a particle and is defined to be:
\begin{quotation}
{\tt double SPINUP(I)}: {\it cosine of the angle between the spin-vector of particle I and the 3- momentum of the decaying particle, specified in the lab frame}
%
%This scheme is neither general nor complete, but is chosen as the best compromise. The main foreseen 
%application is $\tau$s with a specific helicity. Typically a relativistic $\tau-$ ($\tau+$) from a $W-$ ($W+$) 
%has helicity $-1$ ($+1$) (though this might be changed by the boost to the lab frame), so {\tt SPINUP(I)}= $-1$ ($+1$). 
%The use of a floating point number allows for the extension to the non-relativistic case. Unknown or unpolarized 
%particles should be given {\tt SPINUP(I)}=9. The lab frame is the frame in which the four-vectors are specified.
\end{quotation}
This definition of spin for particles in Monte Carlos is unfortunately complicated and not widely applicable.
Its use has been mainly restricted to treating polarized $\tau$ decays.  
%We advocate for abandoning {\tt SPINUP} and defining the spin of a 
%particle by its chirality for fermions or by its helicity for massless vector bosons.  
In this and future work, we propose using chirality as the basis for defining spin for massive or
massless fermions.
Chirality is Lorentz invariant, relevant in weak decays and reduces to helicity for massless
fermions.

Also, polarized splitting functions \cite{stirling96} should be distinguished from helicity splitting functions.
It is first an issue of semantics.  Helicity is the handedness of circular polarization of a 
particle with respect to its momentum.  For massless particles, this is Lorentz invariant, as mentioned
earlier.  Polarized splitting functions instead reference the linear polarization of a
particle with respect to the plane of the splitting.  They are thus not Lorentz invariant,
even for massless particles.  However, polarized splitting functions can be used
to approximate the azimuthal correlations between 
subsequent emissions and the effect 
on the energy distribution of the shower.  Helicity splitting
functions do not have this property; 
however, the azimuthal correlations do of course reappear 
when the shower is matched to matrix elements.  

\subsection{Global Antennae\label{subsec:GlobalAntennae}}

\begin{table}[tp]
\begin{center}{
\begin{tabular}{lr|rrrrrr|rrr}
\toprule 
$\times$ & \hspace*{-2mm}$\frac{1}{y_{ij}y_{jk}}$
& $\frac{1}{y_{ij}}$ & $\frac{1}{y_{jk}}$ & $\frac{y_{jk}}{y_{ij}}$ &
$\frac{y_{ij}}{y_{jk}}$
& $\frac{y_{jk}^2}{y_{ij}}$ & $\frac{y_{ij}^2}{y_{jk}}$ & $1$ & $y_{ij}$ & $y_{jk}$\\[2mm]
\hline
\multicolumn{4}{l}{$q\bar{q}\to qg\bar{q}$}\\
$++\to +++$ & 1 & 0 & 0 & 0 & 0 & 0 & 0 & 0 & 0 & 0\\
$++\to +-+$ & 1 & $-2$ & $-2$ & 1 & 1 & 0 & 0 & 2 & 0 & 0\\
$+-\to ++-$ & 1 & 0 & $-2$ & 0 & 1 & 0 & 0 & 0 & 0 & 0\\
$+-\to +--$ & 1 & $-2$ & 0 & 1 & 0 & 0 & 0 & 0 & 0 & 0\\
\hline
\multicolumn{4}{l}{$qg\to qgg$}\\
$++\to +++$ & 1 & 0 & $-\alpha+1$ & 0 & $2\alpha-2$ & 0 & 0 & 0 & 0 & 0\\
$++\to +-+$ & 1 & $-2$ & $-3$ & 1 & 3 & 0 & $-1$ & 3 & 0 & 0\\
$+-\to ++-$ & 1 & 0 & $-3$ & 0 & 3 & 0 & $-1$ & 0 & 0 & 0\\
$+-\to +--$ & 1 & $-2$ & $-\alpha+1$ & 1 & $2\alpha-2$ & 0 & 0 & 0 & 0 & 0\\
\hline
\multicolumn{4}{l}{$gg\to ggg$}\\
$++\to +++$ & 1 & $-\alpha+1$ & $-\alpha+1$ & $2\alpha-2$ & $2\alpha-2$ & 0 & 0 & 0 & 0 & 0\\
$++\to +-+$ & 1 & $-3$ & $-3$ & 3 & 3 &$-1$ & $-1$ & 3 & 1 & 1\\
$+-\to ++-$ & 1 & $-\alpha+1$ & $-3$ & $2\alpha-2$ & 3 &0 & $-1$ & 0 & 0 & 0\\
$+-\to +--$ & 1 & $-3$ & $-\alpha+1$ & 3 &  $2\alpha-2$ & $-1$ & 0 & 0 & 0 & 0\\
\hline
\multicolumn{4}{l}{$qg\to q\bar{q}'q'$}\\
$++\to ++-$ & 0 & 0 & 0 & 0 & 0 & 0 & $\frac{1}{2}$ & 0 & 0 & 0\\
$++\to +-+$ & 0 & 0 & $\frac{1}{2}$ & 0 & $-1$ & 0 & $\frac{1}{2}$ & 0 & 0 & 0\\
$+-\to ++-$ & 0 & 0 & $\frac{1}{2}$ & 0 & $-1$ & 0 & $\frac{1}{2}$ & 0 & 0 & 0\\
$+-\to +-+$ & 0 & 0 & 0 & 0 & 0 & 0 & $\frac{1}{2}$ & 0 & 0 & 0\\
\hline
\multicolumn{4}{l}{$gg\to g\bar{q}q$}\\
$++\to ++-$ & 0 & 0 & 0 & 0 & 0 & 0 & $\frac{1}{2}$ & 0 & 0 & 0\\
$++\to +-+$ & 0 & 0 & $\frac{1}{2}$ & 0 & $-1$ & 0 & $\frac{1}{2}$ & 0 & 0 & 0\\
$+-\to ++-$ & 0 & 0 & $\frac{1}{2}$ & 0 & $-1$ & 0 & $\frac{1}{2}$ & 0 & 0 & 0\\
$+-\to +-+$ & 0 & 0 & 0 & 0 & 0 & 0 & $\frac{1}{2}$ & 0 & 0 & 0\\
\bottomrule
\end{tabular}}
\caption{Table of coefficients for helicity-dependent global antenna
functions.  By the $C$ and $P$ invariance of QCD, the same expressions
apply with $+\leftrightarrow -$, $q\leftrightarrow \bar{q}$.  All other antennae
are zero.  The parameter $\alpha$ determines the form of the spin-summed
global antennae.  The default choice in \Vc is $\alpha = 0$
which corresponds to the Gehrmann-De Ridder, Gehrmann, and Glover (GGG) spin-summed antennae \cite{GehrmannDeRidder:2005cm}.  
The finite terms are chosen so that the antennae are positive on all of final-state phase space.
\label{tab:glob_antenna_coefficients}}
\end{center}
\end{table}

The forms of the global antennae are found by enforcing several requirements.
Global antennae contain the full soft limit of emitted gluons but neighboring
antennae share a collinear limit of gluons.  To construct the helicity-dependent global antennae, then,
every possible helicity configuration of neighboring antennae must reproduce the
correct collinear limits.  Also the helicity dependent antennae can become negative
over a significant region of phase space.  For the use of the antennae functions as 
probability distributions on phase space they must be positive on all of phase space.

For the unpolarized global antennae, it is a straightforward exercise to incorporate
all constraints to determine the antennae.  We present an example of this 
in~\appRef{app:globalunpol}.  The construction of helicity-dependent global antennae
is more subtle, but we employ the following requirements to simplify the analysis:
\begin{enumerate}
\item Bose-Einstein symmetry.  The antenna functions must be symmetric when gluons of
the same helicity are exchanged.

\item $C$ and $P$ symmetry of QCD.  The expressions for the antennae are unchanged
with $+ \leftrightarrow -$, $q\leftrightarrow \bar{q}$.

\item Neighboring antennae sum to reproduce the full collinear limits.

\item The singular terms of the helicity-dependent global antennae must sum to reproduce
the singular terms of the unpolarized global antennae.

\item Positivity of global antennae.  Because the collinear limits of gluons are constructed from the
sum of neighboring antennae, the antennae are not guaranteed to be positive even in the
singular regions of phase space.  The positivity requirement must be
enforced in the singular as well as non-singular regions of phase space.

\end{enumerate}
A careful accounting of these requirements produces helicity-dependent global antennae that
depend on three arbitrary parameters.  One of these parameters fixes the form of the 
spin-summed or unpolarized antennae, which we call $\alpha$, while the other two are artifacts of the proliferation
of helicity-dependent antennae.  The latter two parameters can consistenly be set to zero, which
we choose to do in the following.  The complete procedure, with fully general
expressions, is described in  \appRef{app:globalhel}. Here, we just give
the forms of the single-parameter antenna functions implemented in
\Vc, which are defined in \tabRef{tab:glob_antenna_coefficients}.

To estimate shower uncertainties due to the ambiguous choice of
non-singular terms, we define a set of MIN and MAX antenna functions 
which are smaller and larger, respectively, over all of phase space
than the default antennae.  For the MAX antennae, the finite terms of
all helicity-dependent antennae are fixed to a large constant value; we
choose to set the constant to be $5.0$.  This is large enough to guarantee
that all antennae are positive on phase space.  MIN antennae are more subtle
because some helicity-dependent antennae cannot be decreased and
remain positive on phase space.  To assuage this, we choose constants
to subtract from those helicity-dependent antennae which are large and positive
enough to allow this. 
(Specifically, we subtract the minimum value on the $2\to 3$
  phase-space of the singular pieces of the given antenna.)
This procedure guarantees that, when summed over helicities,
the MIN antennae are smaller than the default antennae in \Vc.  
We present the choice of MIN antennae
in~\tabRef{tab:glob_antenna_coefficients_MIN}. 

\begin{table}[tp]
\begin{center}{
\begin{tabular}{lr|rrrrrr|r}
\toprule 
$\times$ & \hspace*{-2mm}$\frac{1}{y_{ij}y_{jk}}$
& $\frac{1}{y_{ij}}$ & $\frac{1}{y_{jk}}$ & $\frac{y_{jk}}{y_{ij}}$ &
$\frac{y_{ij}}{y_{jk}}$
& $\frac{y_{jk}^2}{y_{ij}}$ & $\frac{y_{ij}^2}{y_{jk}}$ & $1$ \\[2mm]
\hline
\multicolumn{4}{l}{$q\bar{q}\to qg\bar{q}$}\\
$++\to +++$ & 1 & 0 & 0 & 0 & 0 & 0 & 0 & -4 \\
$++\to +-+$ & 1 & -2 & -2 & 1 & 1 & 0 & 0 & 2 \\
\hline
\multicolumn{4}{l}{$qg\to qgg$}\\
$++\to +++$ & 1 & 0 & 1 & 0 & -2 & 0 & 0 & -3 \\
$++\to +-+$ & 1 & -2 & -3 & 1 & 3 & 0 & -1 & 3\\
\hline
\multicolumn{4}{l}{$gg\to ggg$}\\
$++\to +++$ & 1 & 1 & 1 & -2 & -2 & 0 & 0 & -4 \\
$++\to +-+$ & 1 & -3 & -3 & 3 & 3 &-1 & -1 & 3.7\\
\bottomrule
\end{tabular}}
\caption{Table of coefficients for MIN helicity-dependent global antenna
functions.  In this table, $\alpha$ has been set to 0  which is the
default choice in \Vc.  Only those
antennae with non-zero finite terms are shown.
\label{tab:glob_antenna_coefficients_MIN}}
\end{center}
\end{table}

\subsection{Sector Antennae}

The helicity-dependent sector antennae are defined by reproducing the appropriate Altarelli-Parisi splitting functions
as the emitted particle becomes collinear with respect to either of the initial particles.
This requirement uniquely fixes the singular components of all sector antennae
but still allows for freedom of the choice of non-singular terms in the antennae.  The non-singular
terms can be chosen so that the sector antennae reproduce matrix elements for particular processes, for example,
which was done in~\cite{Larkoski:2009ah}.  Positivity of the sector antenna functions is guaranteed in the singular regions
because the antennae reproduce the universal Altarelli-Parisi functions.  However, for some antennae,
non-singular pieces must be added to keep the antennae positive in the non-singular regions of phase space.

Defining the antenna functions by a ratio of matrix elements is one prescription for choosing the non-singular terms
that are necessary to enforce positivity of the antenna on all of phase space.
Our prescription for the choice of non-singular terms for the sector antennae is to add only the minimal
terms necessary.  For antennae whose singular terms are positive on all of phase space, we choose
to set the non-singular terms to $0$.  For those antennae which require the addition of non-singular terms
for positivity, we choose to add constants where possible and only include higher order terms in $y_{ij}$ and $y_{jk}$
if necessary for simplicity.  An example of the construction of sector antennae from the collinear limits and positivity
is given in~\appRef{app:sector_ant} and the coefficients of the terms in the sector antennae are given 
in~\tabRef{tab:sec_antenna_coefficients}.

To estimate shower uncertainties due to the ambiguous choice of
non-singular terms, we define a set of MIN and MAX antenna functions,
as in the global shower case.  For simplicity, the finite terms for the 
sector MIN and MAX antennae are chosen to be 
the same as those in the global case.

In the \Vc code, the sector antennae are derived from the global
antennae.  Note from~\tabRef{tab:glob_antenna_coefficients}  
and \tabRef{tab:sec_antenna_coefficients} that much of the structure
of the sector antennae is captured  
by the global antennae if $\alpha=1$.  To construct a sector antenna,
the corresponding global antenna 
with the same helicity  and flavor structure is evaluated with
$\alpha=1$ and the missing terms added to  
recover the full sector antenna.  The precise relationship between the sector ($\bar{a}^{\text{sct}}$) and global ($\bar{a}^{\text{gl}}$)
antennae for $\alpha=1$ for gluon emission is:
\begin{eqnarray}
\bar{a}^{\text{sct}}_{j/IK}(y_{ij},y_{jk}) = \bar{a}^{\text{gl}}_{j/IK}(y_{ij},y_{jk}) &+ &\delta_{Ig}\delta_{h_K h_k}\left\{\delta_{h_I h_i}\delta_{h_I h_j} \left(  \frac{1+y_{jk}+y_{jk}^2}{y_{ij}}
\right) \right.\nonumber \\
&+&\left.\delta_{h_I h_j}\left( \frac{1}{y_{ij}(1-y_{jk})} - \frac{1+y_{jk}+y_{jk}^2}{y_{ij}}\right) \nonumber
\right\} \nonumber \\
&+ &\delta_{Kg}\delta_{h_I h_i}\left\{\delta_{h_I h_j} \delta_{h_K h_k}\left(  \frac{1+y_{ij}+y_{ij}^2}{y_{jk}}
\right) \right.\nonumber \\
&+&\left. \delta_{h_K h_j}\left( \frac{1}{y_{jk}(1-y_{ij})} - \frac{1+y_{ij}+y_{ij}^2}{y_{jk}}\right) \nonumber
\right\} \ .
\end{eqnarray}
Here, $\delta_{Ig}$ is one if $I$ is a gluon and zero otherwise and $\delta_{h_i h_j}$ is one if the helicity
of particles $i$ and $j$ are the same and zero otherwise.
For antennae with gluons splitting to quarks, the sector antennae are twice the global antennae.

\begin{table}[tp!]
\begin{center}{
\begin{tabular}{lr|rrrrrr|rr|rrr}
\toprule 
$\times$ & \hspace*{-2mm}$\frac{1}{y_{ij}y_{jk}}$
& $\frac{1}{y_{ij}}$ & $\frac{1}{y_{jk}}$ & $\frac{y_{jk}}{y_{ij}}$ &
$\frac{y_{ij}}{y_{jk}}$
& $\frac{y_{jk}^2}{y_{ij}}$ & $\frac{y_{ij}^2}{y_{jk}}$ & $\frac{1}{y_{jk}(1-y_{ij})}$&
$\frac{1}{y_{ij}(1-y_{jk})}$ & $1$ & $y_{ij}$ & $y_{jk}$\\[2mm]
\hline
\multicolumn{4}{l}{$q\bar{q}\to qg\bar{q}$}\\
$++\to +++$ & 1 & 0 & 0 & 0 & 0 & 0 & 0 &0&0& 0 & 0 & 0\\
$++\to +-+$ & 1 & -2 & -2 & 1 & 1 & 0 & 0 &0&0& 2 & 0 & 0\\
$+-\to ++-$ & 1 & 0 & -2 & 0 & 1 & 0 & 0 &0&0& 0 & 0 & 0\\
$+-\to +--$ & 1 & -2 & 0 & 1 & 0 & 0 & 0 &0&0& 0 & 0 & 0\\
\hline
\multicolumn{4}{l}{$qg\to qgg$}\\
$++\to +++$ & 1 & 0 & 0 & 0 & 0 & 0 & 0 &1&0& 0 & 0 & 0\\
$++\to +-+$ & 1 & -2 & -3 & 1 & 3 & 0 & -1 &0&0& 3 & 0 & 0\\
$++\to ++-$ & 0 & 0 & -1 & 0 & -1 & 0 & -1 &1&0& 0 & 0 & 0\\
$+-\to ++-$ & 1 & 0 & -3 & 0 & 3 & 0 & -1 &0&0& 0 & 0 & 0\\
$+-\to +--$ & 1 & -2 & 0 & 1 & 0 & 0 & 0 &1&0& 0 & 0 & 0\\
$+-\to +-+$ & 0 & 0 & -1 & 0 & -1 & 0 & -1 &1&0& 0 & 0 & 0\\
\hline
\multicolumn{4}{l}{$gg\to ggg$}\\
$++\to +++$ & 1 & 0 & 0 & 0 & 0 & 0 & 0 &1&1& 0 & 0 & 0\\
$++\to +-+$ & 1 & -3 & -3 & 3 & 3 &-1 & -1 &0&0& 3 & 1 & 1\\
$++\to ++-$ & 0 & 0 & -1 & 0 & -1 &0 & -1 &1&0& 0 & 0 & 0\\
$++\to -++$ & 0 & -1 & 0 & -1 & 0 &-1 & 0 &0&1& 0 & 0 & 0\\
$+-\to ++-$ & 1 & 0 & -3 & 0 & 3 &0 & -1 &0&1& 0 & 0 & 0\\
$+-\to +--$ & 1 & -3 & 0 & 3 & 0 &-1 & 0 &1&0& 0 & 0 & 0\\
$+-\to +-+$ & 0 & 0 & -1 & 0 & -1 &0 & -1 &1&0& 0 & 0 & 0\\
$+-\to -+-$ & 0 & -1 & 0 & -1 & 0 &-1 & 0 &0&1& 0 & 0 & 0\\
\hline
\multicolumn{4}{l}{$qg\to q\bar{q}'q'$}\\
$++\to ++-$ & 0 & 0 & 0 & 0 & 0 & 0 & 1 &0&0& 0 & 0 & 0\\
$++\to +-+$ & 0 & 0 & 1 & 0 & -2 & 0 & 1 &0&0& 0 & 0 & 0\\
$+-\to ++-$ & 0 & 0 & 1 & 0 & -2 & 0 & 1 &0&0& 0 & 0 & 0\\
$+-\to +-+$ & 0 & 0 & 0 & 0 & 0 & 0 & 1 &0&0& 0 & 0 & 0\\
\hline
\multicolumn{4}{l}{$gg\to g\bar{q}q$}\\
$++\to ++-$ & 0 & 0 & 0 & 0 & 0 & 0 & 1 &0&0& 0 & 0 & 0\\
$++\to +-+$ & 0 & 0 & 1 & 0 & -2 & 0 & 1 &0&0& 0 & 0 & 0\\
$+-\to ++-$ & 0 & 0 & 1 & 0 & -2 & 0 & 1 &0&0& 0 & 0 & 0\\
$+-\to +-+$ & 0 & 0 & 0 & 0 & 0 & 0 & 1 &0&0& 0 & 0 & 0\\
\bottomrule
\end{tabular}}
\caption{Table of coefficients for helicity-dependent sector antenna
functions.  By the $C$ and $P$ invariance of QCD, the same expressions
apply with $+\leftrightarrow -$, $q\leftrightarrow \bar{q}$.  All other antennae
are zero.  These are the default assignments in \Vc.  The finite terms are
chosen so that the antennae are positive on all of final-state phase space.
\label{tab:sec_antenna_coefficients}}
\end{center}
\end{table}

\section{Results \label{sec:results}}

\subsection{Comparison to Matrix Elements}

In order to examine the quality of the approximation furnished 
by a shower based on the antennae derived above, 
independently of the shower code, we follow the approach used 
for global and sector unpolarized antennae in \cite{Skands:2009tb,Giele:2011cb,GehrmannDeRidder:2011dm,LopezVillarejo:2011ap}.  We use 
\textsc{Rambo} \cite{Kleiss:1985gy} (an implementation of which
has been included in \Vc) to generate uniformly distributed 4-, 5-,
and 6-parton phase-space points. At 
each phase-space point, we use \Mg v.~4.4.26~\cite{Alwall:2007st} 
and the HELAS libraries~\cite{Murayama:1992gi}
to evaluate the leading-color, helicity-dependent matrix element.  As
with matching, 
the \Mg code has been modified to extract individual helicity configurations
and color structures.

For each phase-space point and helicity configuration, the corresponding antenna shower approximation to the matrix element is then computed.
This is done by using a clustering algorithm that contains the exact inverse of the default \Vc $2\to3$ kinematics map \cite{Giele:2007di}.
This $3\to 2$ clustering procedure is continued until the desired matched order is reached.  At each step in the kinematic clustering procedure, the 
antennae corresponding to all possible intermediate spins that could have been generated by the helicity-dependent shower are summed over.
To match the global shower to matrix elements requires summing over all possible kinematic clustering histories.  The sector shower,
by contrast, has a unique kinematic history.  To determine which sector is clustered in each step, a partitioning variable must be used. Our
default sector decomposition prescription is based on the variable $Q_{\mrm{s}_j}^{2}$, defined in \cite{LopezVillarejo:2011ap}. 
The three-parton configuration with the smallest value of $Q_{\mrm{s}_j}^{2}$ gets clustered. 
This procedure produces the shower approximation to the matrix element as a nested product of 
helicity-dependent antenna functions.

\begin{figure}[tp]
\centering
\includegraphics*[scale=0.75]{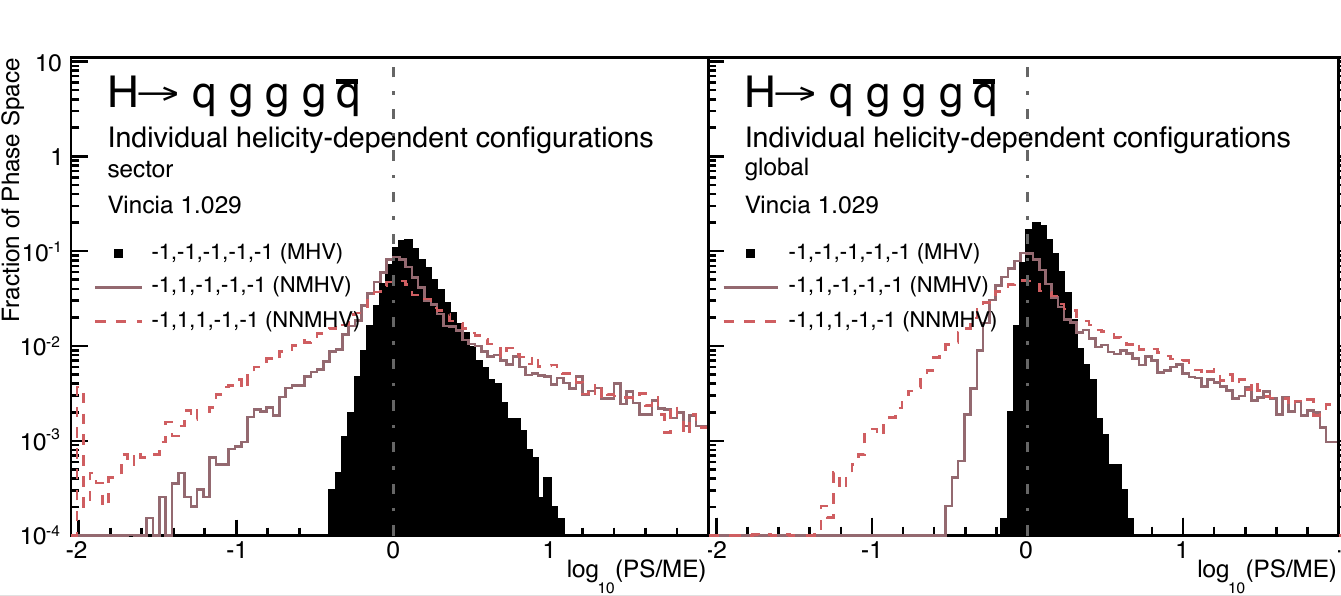}
\caption{Accuracy of individual configurations in the shower
  approximation compared to helicity-dependent LO matrix elements for
  $H\to q g g g \bar{q}$. Distributions of
  $\log_{10}(\mrm{PS}/\mrm{ME})$ in a flat phase-space  
  scan, normalized to unity.}
\label{spin_indivSpinConf_Htoqgggqbar}
\end{figure}

\begin{figure}[tp]
\centering
\includegraphics*[scale=0.75]{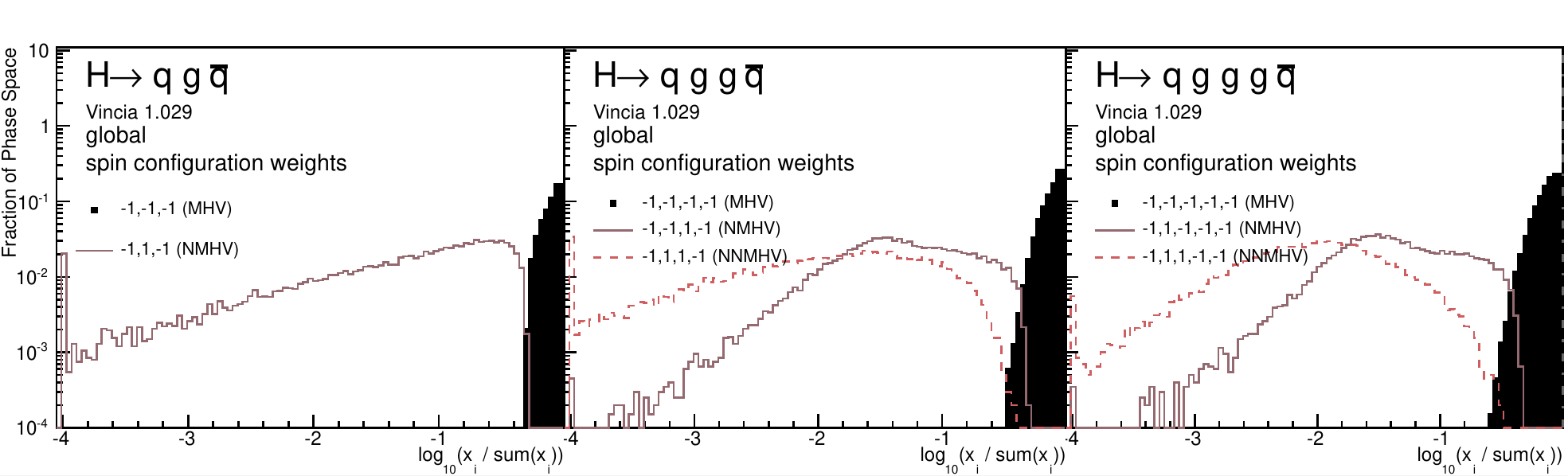}
\includegraphics*[scale=0.75]{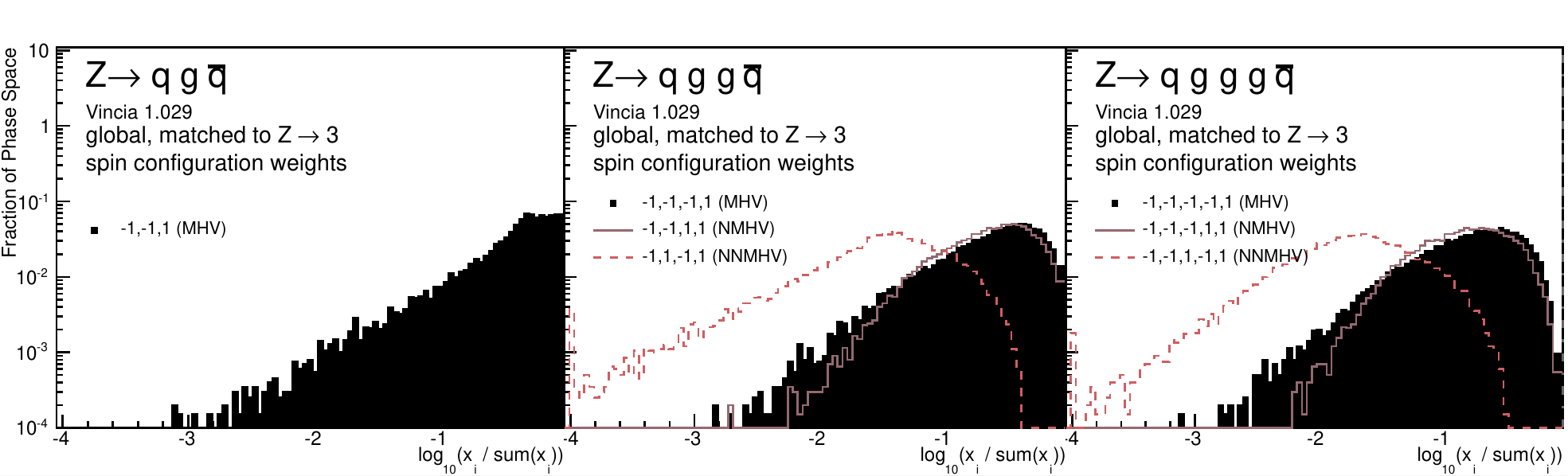}
\caption{Relative weight of some specific  helicity configurations in the global shower approximation to LO matrix elements for $H\to q\bar{q}\,+\,$gluons (above) and $Z\to q\bar{q}\,+\,$gluons (below). The sector shower displays basically the same structure, in particular the same hierarchy MHV, NMHV, NNMHV. The sum $\sum{x_i}$ runs over all helicity configurations with the same helicities for  $q\bar{q}$ and includes, in some cases, configurations that are not being plotted. Distributions of $\log_{10}(\mrm{PS}/\mrm{ME})$ in a flat phase-space scan, normalized to unity.}
\label{spinweights_global_HandZ}
\end{figure}

To compare the shower to matrix elements we will consider $Z$ and $H$ decays to quarks with additional radiation.  We begin by comparing directly the
helicity matrix elements to the helicity-shower approximation.  In the comparison, we will organize the helicity configurations by their complexity.  We refer
to processes with the maximum number of like helicities with the standard name of maximally helicity violating (MHV).  Processes with one spin flip with
respect to MHV we refer to as next-to-MHV (NMHV), and similarly for more complex spin configurations.  We will see that the helicity-dependent shower approximates the
MHV helicity matrix elements very well, and the accuracy of the shower decreases as the helicity structure becomes more complicated.  However, NMHV and higher 
helicity configurations are subdominant contributions to the spin-summed process, in general.  Thus, when summing over spins, we expect the helicity shower
to have comparable accuracy to the spin-summed matrix element as for MHV configurations.

In \figRef{spin_indivSpinConf_Htoqgggqbar}, we compare the parton shower approximation to the matrix element for the process $H \to q g g g \bar{q}$  for different spin configurations
of the gluons.  As expected, the MHV configuration is best approximated by the helicity shower.  Also, note that the global shower is more accurate than the sector shower,
for the same spin configuration.  Note, however, that flat phase space is unphysical, and the effective accuracy of the shower will actually be significantly better
for realistic phase-space weighting.

It is interesting to compare the relative weight of the spin configurations generated by the helicity shower.   In \figRef{spinweights_global_HandZ} we plot the ratio of the matrix element approximation
from the global helicity shower to the spin-summed matrix element approximation for both $H$ and $Z$ decay processes.  The corresponding plots for the sector shower are similar.
  As expected, the MHV-type matrix elements are most of the spin-summed result.
This is most evident in the plot for the $H$ decay.  Because the $H$ is a scalar, the helicity of the quark decay products must be the same and so the MHV configuration consists of all quarks and gluons
with the same helicity.  By contrast, because the $Z$ is a vector, the helicities of the quarks are different and so the MHV configuration already has one spin flipped.  

\begin{figure}[tp]
\centering
\includegraphics*[scale=0.75]{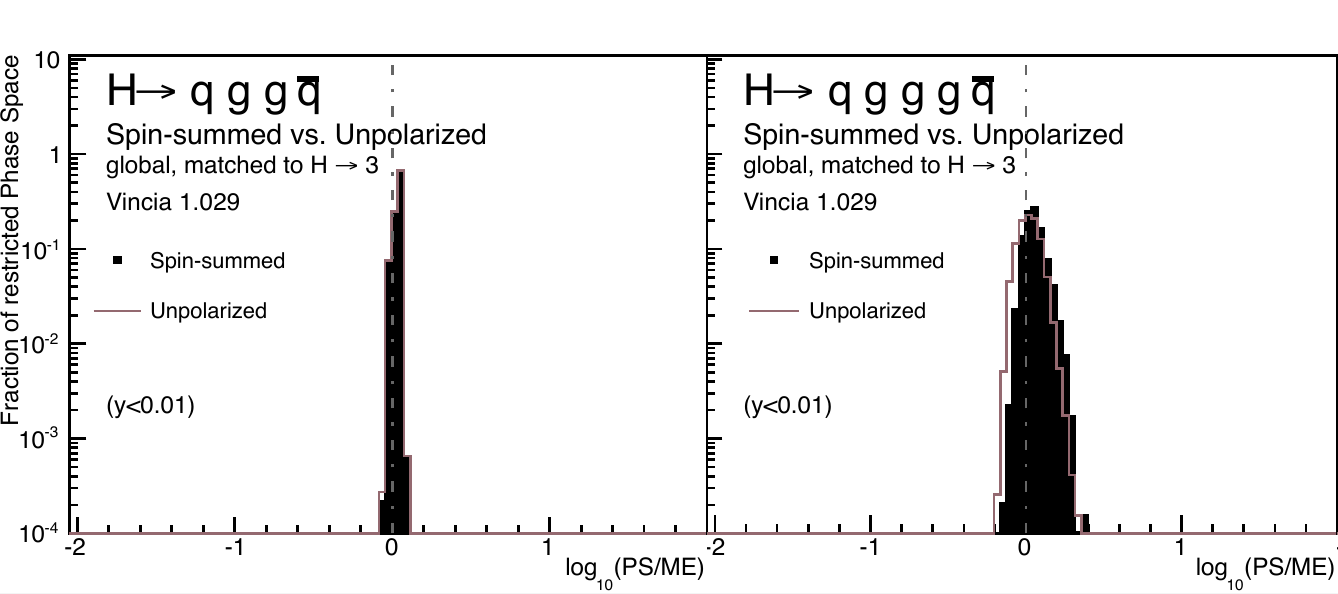}
\includegraphics*[scale=0.75]{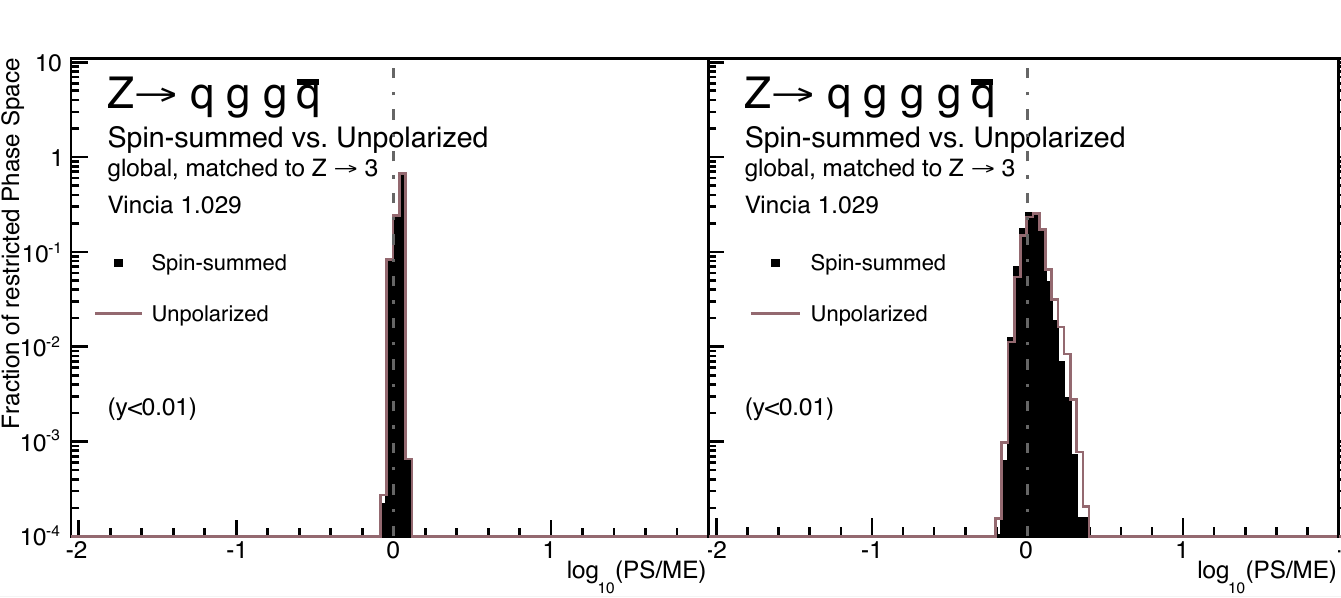}
\caption{Global showers. Spin-summed helicity-dependent and
  unpolarized shower approximations compared to LO matrix elements for $H\to q\bar{q}\,+\,$gluons (above) 
and  $Z\to q\bar{q}\,+\,$gluons (below). Distributions of
  $\log_{10}(\mrm{PS}/\mrm{ME})$ in a flat phase-space 
  scan, normalized to unity, with hard configurations excluded.
\label{spinUnpolVsPol_global_HandZ}}
\end{figure}

\begin{figure}[tp]
\centering
\includegraphics*[scale=0.75]{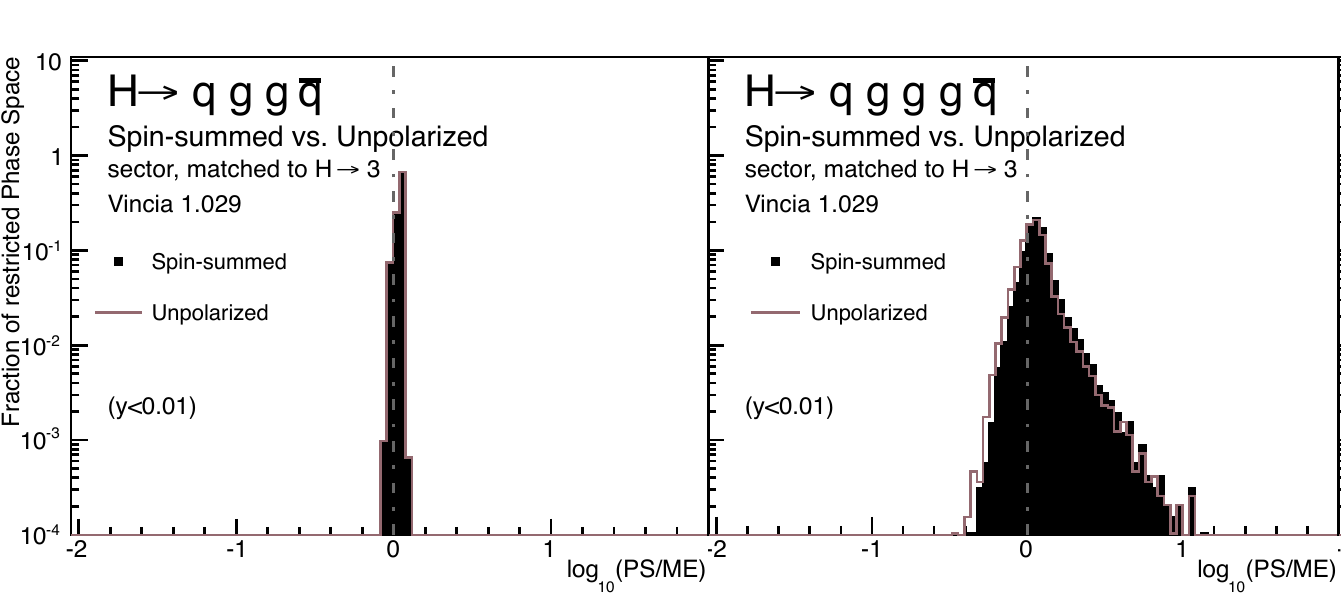}
\includegraphics*[scale=0.75]{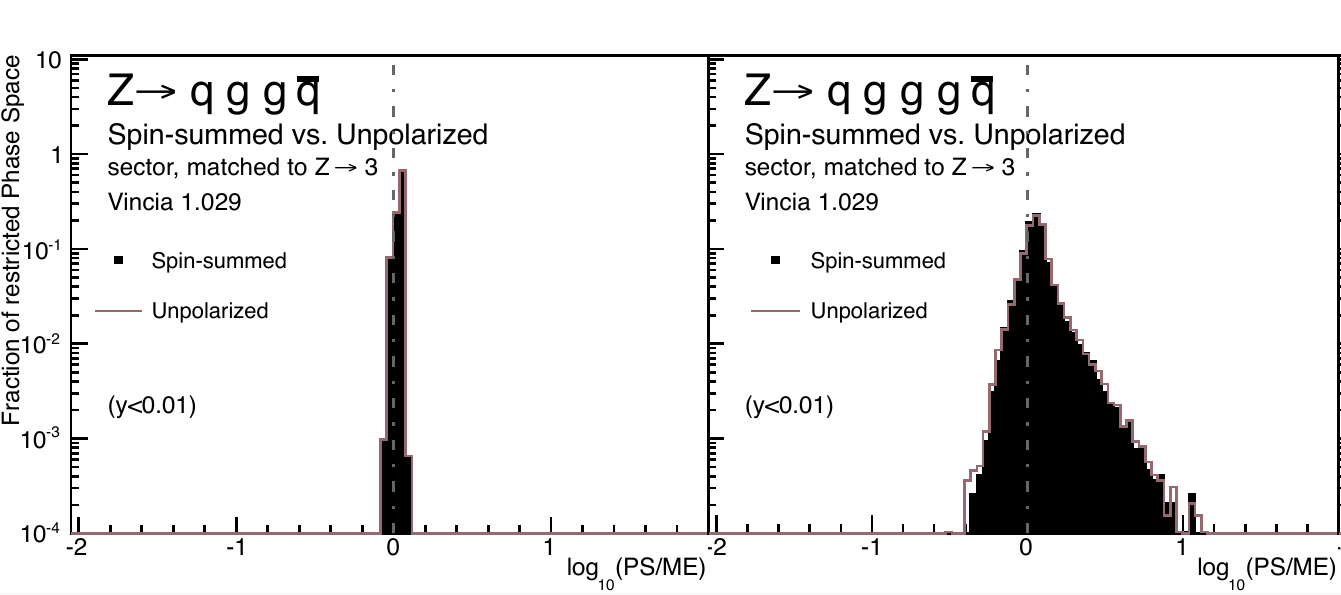}
\caption{Sector showers. Spin-summed helicity-dependent and
  unpolarized shower approximations compared to LO matrix elements for $H\to q\bar{q}\,+\,$gluons (above) 
and  $Z\to q\bar{q}\,+\,$gluons (below). Distributions of
  $\log_{10}(\mrm{PS}/\mrm{ME})$ in a flat phase-space 
  scan, normalized to unity, with hard configurations excluded.}
\label{spinUnpolVsPol_sector_HandZ}
\end{figure}

We now compare spin-summed matrix elements to summing over spin configurations in the helicity shower and to an unpolarized antenna shower as in \cite{Giele:2011cb, LopezVillarejo:2011ap}.  
As mentioned earlier, we expect some level of accuracy improvement with the spin-summed helicity shower as compared to the unpolarized shower.  In the helicity shower, some spin
configurations are not allowed to contribute, while the unpolarized shower gives equal weight to every possible spin configuration.  To compare the two approaches, we will focus on the
singular behavior of the shower, as the non-singular terms are arbitrary anyway.  To do this, we will demand that at least one pair of adjacent partons has a small invariant; namely, we require that
$y_{ij}<0.01$ for neighbors $i$ and $j$.  The ratio of the parton shower approximation to the matrix elements are plotted in \figRef{spinUnpolVsPol_global_HandZ} for the global shower and 
\figRef{spinUnpolVsPol_sector_HandZ} for the sector shower.  Note that there is a small decrease in the width of the distributions for the spin-summed helicity shower with respect to the unpolarized shower,
especially at higher multiplicities.  

\begin{figure}[t]
\centering
\includegraphics*[scale=0.75]{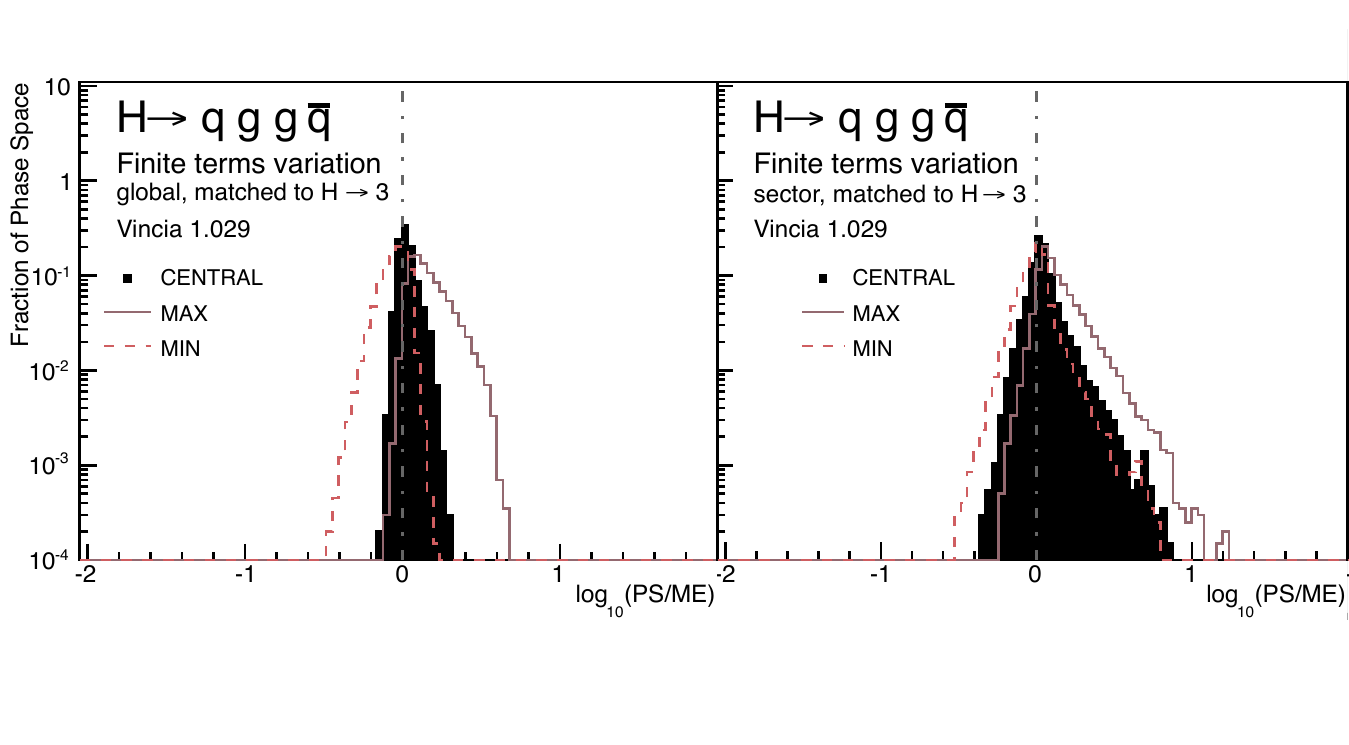}
\includegraphics*[scale=0.75]{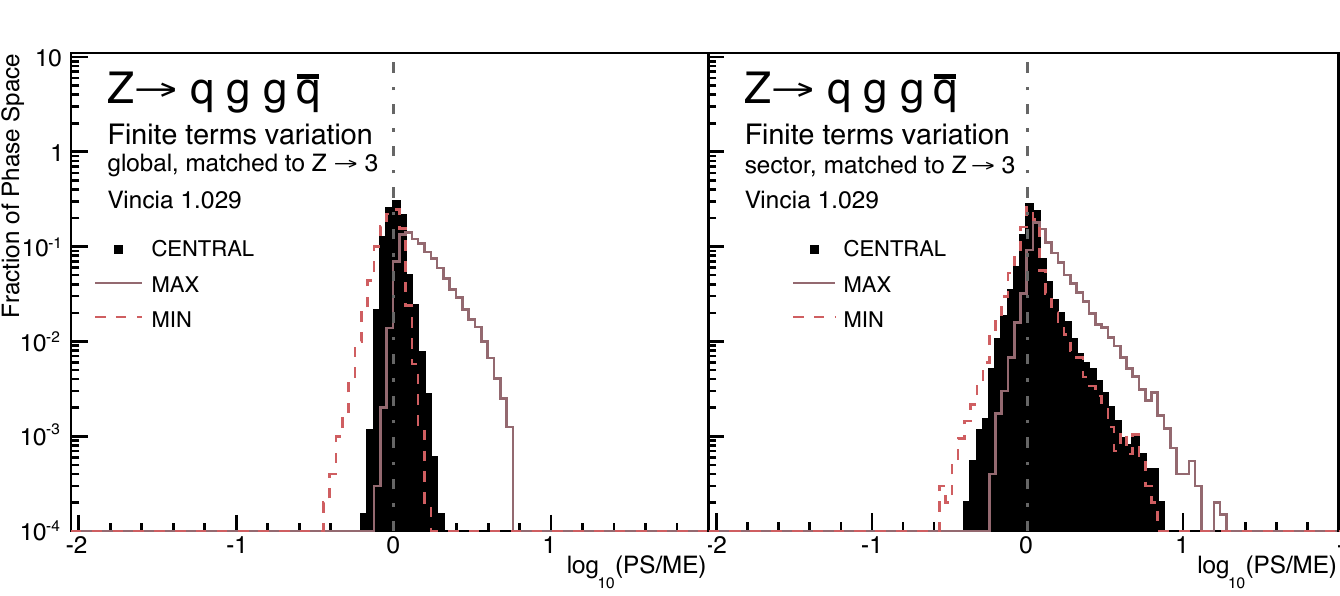}
\caption{Central, MIN and MAX variations of the antennae for the global and sector shower approximations to LO matrix elements for $H\to q\bar{q}\,+\,$gluons and $Z\to q\bar{q}\,+\,$gluons. Distributions of
  $\log_{10}(\mrm{PS}/\mrm{ME})$ in a flat phase-space 
  scan, normalized to unity.}
\label{spinUnpolVsPol_MINandMAXvariations_HandZ}
\end{figure}

However, we do not expect that the helicity-dependent shower is more accurate than the unpolarized shower when considering matrix elements with gluon splitting to quarks. From 
\tabRef{tab:glob_antenna_coefficients} and
\tabRef{tab:sec_antenna_coefficients}, the gluon splitting antennae
reproduce the unpolarized antennae both by summing over the final
spins  
for a given initial spin configuration as well as summing over the initial spins for a final spin configuration.  This implies that, for example, 
the approximation to the spin-summed matrix element for the process
$H\to q \bar{q} q' \bar{q}' $ is exactly the same in the helicity
shower as the unpolarized shower 
(up to non-singular terms). We therefore do not include
comparisons between the two. 

It is also useful to see the dependence of the accuracy of the shower on the choice of arbitrary finite terms in the antennae.  In \figRef{spinUnpolVsPol_MINandMAXvariations_HandZ}
we plot the spin-summed helicity shower with the default, MIN and MAX definitions of the non-singular terms in the antennae from \secRef{sec:formalism}.  Even with these rather extreme
choices for the finite terms in the antennae (especially for MAX), the
shower still gives a good approximation to the matrix elements.  As
the multiplicity increases, the finite terms become less important. 

\begin{figure}[t]
\centering
\includegraphics*[scale=0.33]{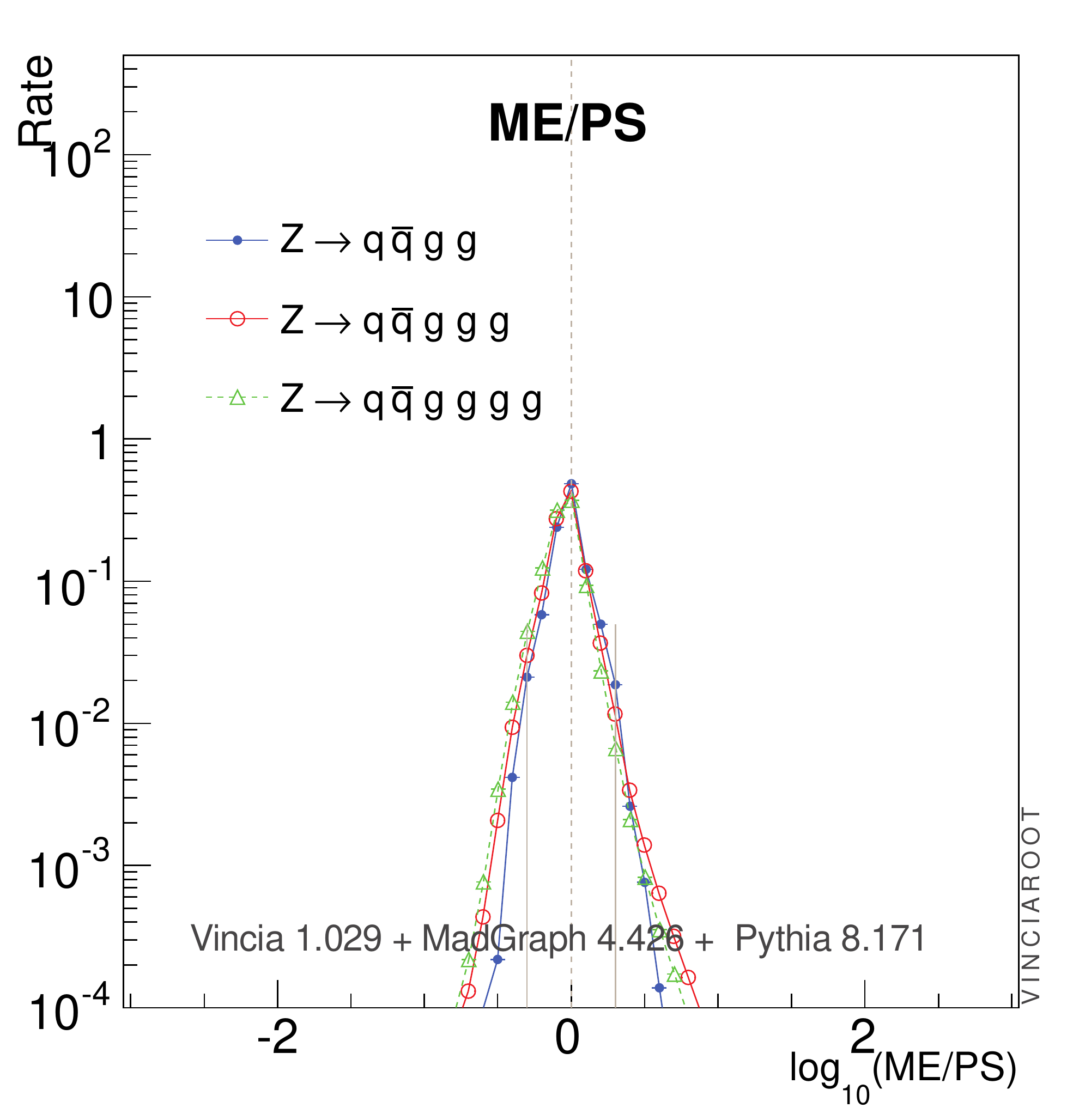}
\includegraphics*[scale=0.33]{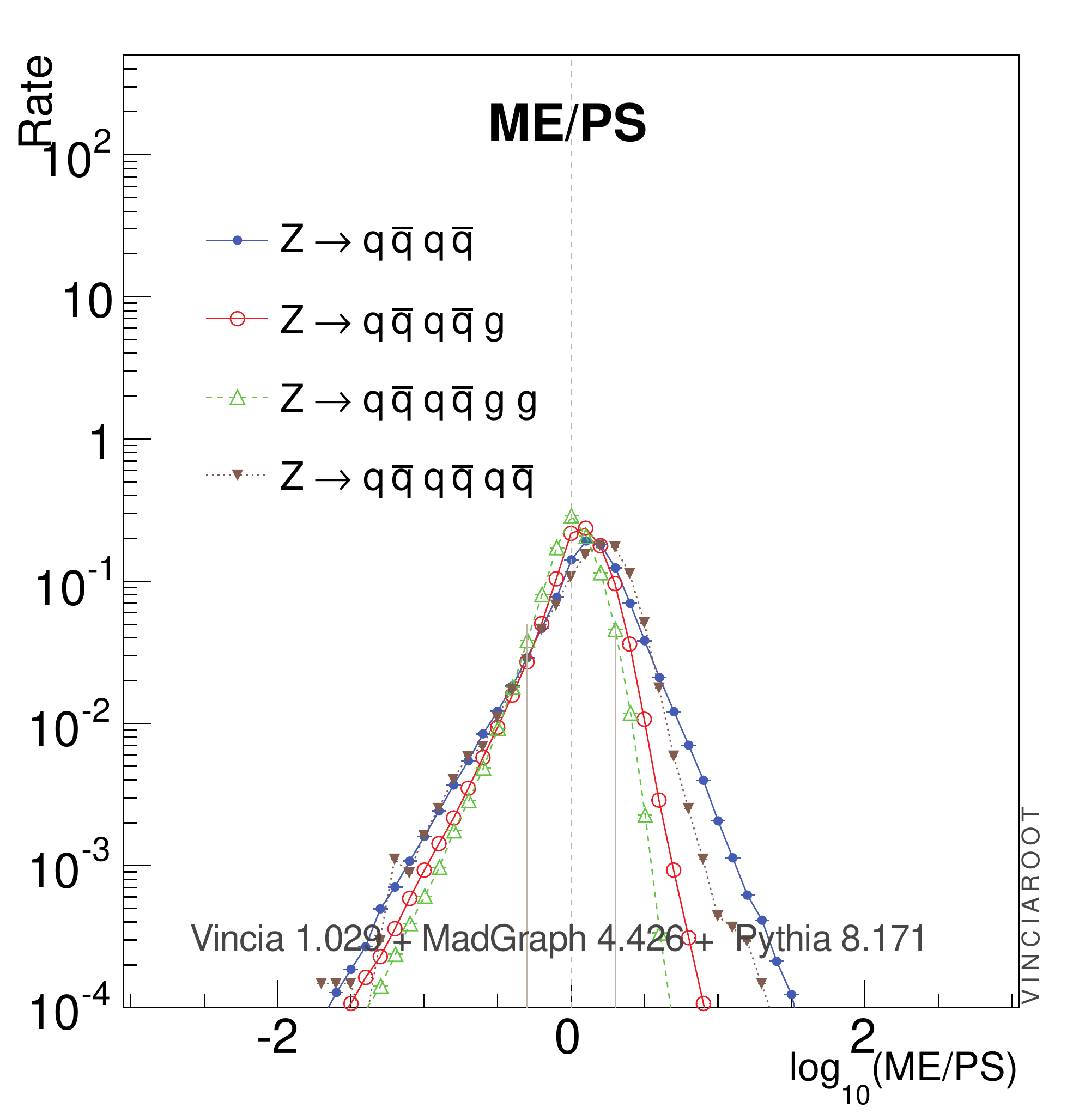}
\caption{Global showers. 
  Distributions of the ME/PS correction factors in actual VINCIA runs,
  for decays of unpolarized $Z$ bosons to massless quarks, using
  helicity-dependent antenna functions. 
{\sl Left:} correction factors for gluon
  emission. 
  {\sl Right:} correction factors for events involving
  $g\to q\bar{q}$ splittings. \label{fig:showerWeights}}
\end{figure}

The previous plots served to illustrate the behavior of the
shower expansions, starting from 3 partons and comparing 
to the tree-level matrix element result at each successive
multiplicity. However, 
in the GKS matching algorithm
implemented in \Vc, the matrix-element corrections are actually
performed sequentially, order by order. That is, the correction to 4 partons is
applied before the evolution goes on to 5 partons, and so on. 
In that context, 
what is relevant for, say, the 6-parton
correction factor is therefore not the pure shower expansion but
rather the approximation obtained from a single branching step starting from
the 5-parton matrix element. Moreover, most of the previous plots 
focused on the shower-dominated regions of phase-space.
 In real life, events will be obtained with a continuous distribution 
 of scales, from hard to soft. 
To illustrate the distribution of correction
factors in actual VINCIA runs, without any phase-space cuts (apart
from the hadronization scale), we make use of the fact that VINCIA 
stores several internal diagnostics histograms
during running, when the verbosity parameter
\ttt{Vincia:verbose} is set to values $\ge 2$. These make use of
PYTHIA's simple histogramming utility and can be printed at the end of
a run by invoking the command \ttt{VinciaShower::printHistos()}. Part 
of these diagnostics histograms contain the ME/PS weight ratios for
both trial and accepted branchings. The latter accurately reflects the
distribution of ME/PS correction factors for each physical branching
that occurs in the evolution. Note, though, that the ratio is here
inverted, from PS/ME to ME/PS; above, we were interested to know
whether the shower over- or under-counted the matrix element. For GKS
matching, we are interested in the size of the correction factor,
which is proportional to ME/PS.

\FigRef{fig:showerWeights} shows a compilation of such plots, for 
$Z\to 4$, $5$, and $6$ partons, using the default global helicity-dependent
showers. The left-hand pane shows
gluon-emission distributions, the three curves representing $Z\to
q\bar{q}gg$, $Z\to q\bar{q}ggg$, and $Z\to q\bar{q}gggg$,
respectively. The central dashed line represents perfect agreement
(the matrix-element correction factor is unity), while
the two solid lines represent a factor two deviation in
each direction. Despite the fact that we are now including hard as well
as soft branchings and that the matching factors now also include
components designed to absorb the subleading-color
corrections~\cite{Giele:2011cb},  
the distributions are still quite narrow. Importantly, we do not observe
any substantial degradation of the correction factor with
multiplicity, suggesting that the GKS matching strategy is quite
stable.

In the right-hand pane of \figRef{fig:showerWeights}, we show the equivalent
distributions for events involving $g\to q\bar{q}$
splittings. (In absolute terms, these events are of course 
less frequent than the gluon-emission ones, but we here normalize
all plots to unity.) As expected, the distributions are broader,
reflecting the fact that the uncorrected cascade is less precise for
this type of branchings, due to the less singular nature of the $g\to
q\bar{q}$ antenna functions. 
The consequence of this is that relatively large trial
overestimates need to be used for $g\to q\bar{q}$ splittings in order
that the tail of large corrections does not lead to unitarity
violations.  Nevertheless, the method appears to remain stable even after multiple
$g\to q\bar{q}$ splittings (the dotted curve shows the comparison to
the $Z\to qqq\bar{q}\bar{q}\bar{q}$ matrix element).

The effect of the GKS matrix-element corrections is to transform the
distributions in \figRef{fig:showerWeights} back to delta functions
(corrected PS = ME) at each order. In particular 
the amount and distribution of $g\to q\bar{q}$ splittings in the matrix-element 
corrected cascade should thus be substantially more accurate than
would be the case in the pure shower.

\subsection{Speed}

A central point of the helicity-based approach presented here 
is that high computational speeds can 
be obtained, even when including matching to quite large partonic
multiplicities. There are essentially three important reasons for
this:
\begin{itemize}
\item The
  initialization time is essentially zero.
In the GKS matching sceme~\cite{Giele:2011cb}, 
only the Born-level cross section needs to be precomputed, and only a
Born-level fixed-order phase-space generator needs to be
initialized, resulting in essentially vanishing initialization times
(of order fractions of a second). This is in contrast to 
slicing-based strategies like
L-CKKW~\cite{Catani:2001cc,Lonnblad:2001iq}, MLM~\cite{Mangano:2006rw}, and
others~\cite{Hamilton:2010wh,Lonnblad:2012ng}  
for which the inclusive 
cross section for each matched multiplicity must be precomputed and a
corresponding $n$-parton phase-space generator initialized (``warmed
up'') before event generation can begin. 
\item In all (unweighted) fixed-order calculations, and consequently
  also in slicing-based matching strategies, one faces the problem
  that QCD amplitudes beyond the first few partons have quite
  complicated structures in phase space. This means that 
  even fairly clever multi-channel strategies have a hard time
  achieving high efficiency over all of it. In GKS, 
  this problem is circumvented by generating the
  phase space by a (trial) shower algorithm, which is
  both algorithmically fast and is guaranteed to get at
  least the leading QCD singularity structures
  right\footnote{A related type of phase-space generator is embodied
    by the SARGE algorithm~\cite{Draggiotis:2000gm}, and there are
    also similarities 
    with the forward-branching scheme proposed in \cite{Giele:2011tm}.}. Since
  those structures give the 
  largest contributions, the fact that the trials are less
  efficient for hard radiation has relatively little impact on the
  overall efficiency\footnote{As long as all of phase-space is covered and
  the trials remain overestimates over all of it, something which we
  have paid particular attention to in \Vc, see
  \cite{Giele:2011cb}.}.
  Combining this with
  the clean properties of the antenna 
  phase-space factorization and with matching at the preceding orders,
  the trial phase-space population at any given parton multiplicity
  is already very close to the correct one, and identical to it in the
  leading singular limits, producing the equivalent of very high
  matching-and-unweighting efficiencies. 
\item Finally, the addition of helicity dependence to the trial
  generation in this paper allows us to match to only a single
  helicity amplitude at a time, at each multiplicity. This gives a
  further speed gain relative to the older
  approach~\cite{Giele:2011cb} in which one 
  had to sum over all helicity configurations at each order. In
  addition, the MHV-type
helicity configurations tend to give the dominant contribution to the spin-summed matrix element.  MHV amplitudes are also those
best described by the shower because they contain the maximum number
of soft and collinear singularities.  
\end{itemize}

\begin{figure}[t]
\centering\vspace*{-10mm}
\includegraphics*[scale=0.45]{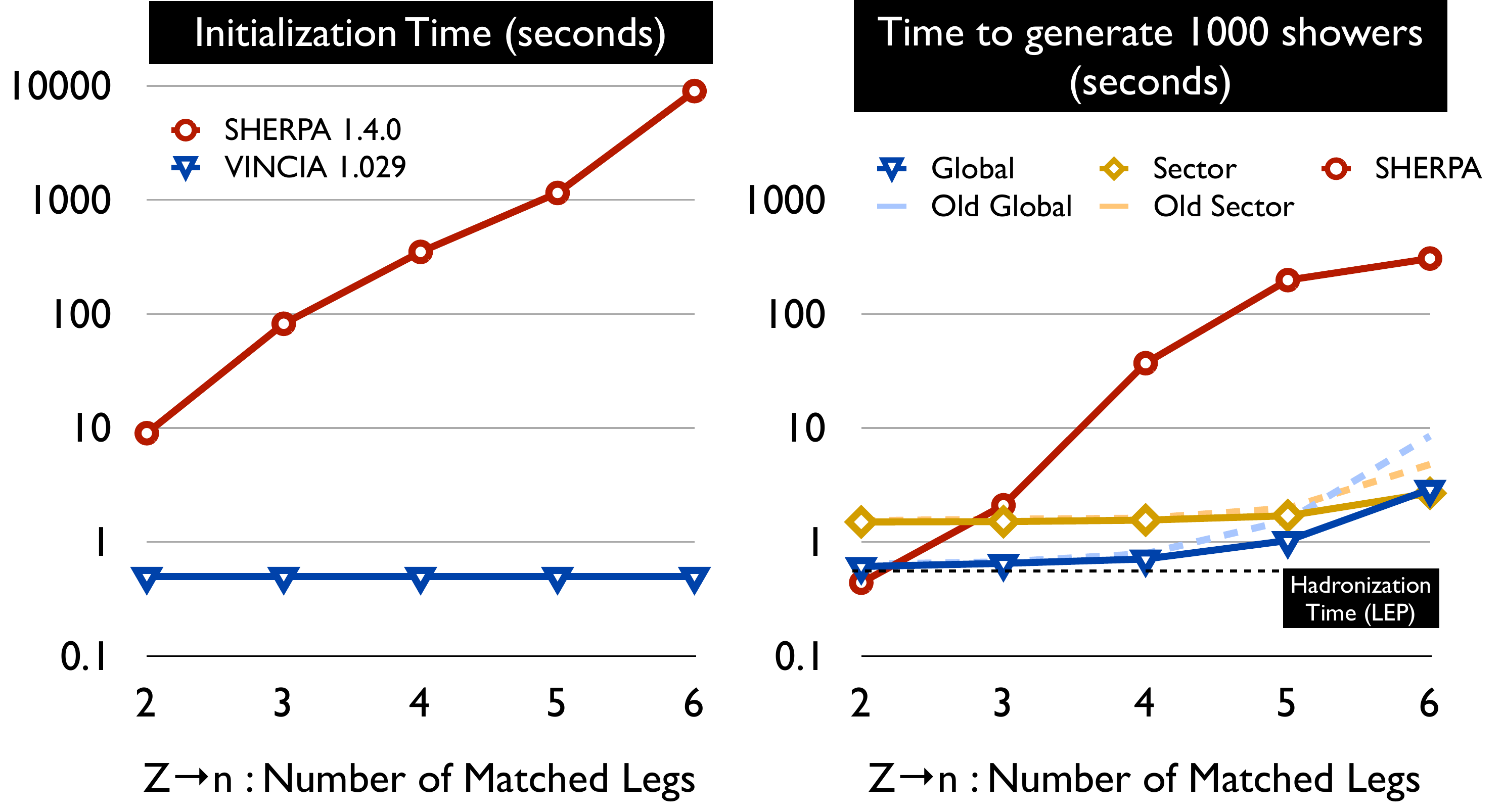}
\caption{Comparison of computation speeds between SHERPA version
  1.4.0~\cite{Gleisberg:2008ta} 
  and \Vc~1.029 + \Py~8.171, as a function of the number of legs
  that are matched to matrix elements, for hadronic $Z$ decays. 
{\sl Left:} initialization time (to precompute cross sections, warm up
phase-space grids, etc, before event generation). {\sl Right:} time to
generate 1000 
parton-level showered events (not including hadronization), with
\Vc's global and sector 
showers shown separately, with and without (``old'') helicity dependence. For
comparison, the average
time it takes to hadronize such events with \Py's string
hadronization model~\cite{Andersson:1998tv} is shown as a dashed
horizontal line.
Further details on the setup used for these runs are given in the
text.
\label{fig:speed}}
\end{figure}
The speed of the old (helicity-independent) \Vc algorithm was
examined in \cite{LopezVillarejo:2011ap}, for the process of $Z$ decay
to quarks plus showers, and was there compared to
SHERPA~\cite{Gleisberg:2008ta}, 
as an example of a slicing-based multileg matching
implementation. In \figRef{fig:speed}, we repeat this comparison,
including now the helicity-dependent \Vc implementations as well. 
Needless to say, other factors play in when comparing
two completely different programs such as SHERPA and \Vc + \Py. 
We do not attempt to account fully for differences in code structures
and optimizations here, so the absolute values
shown in \figRef{fig:speed} should 
not be taken too seriously. Nonetheless, we may take the
results obtained with SHERPA as representative of the scaling
exhibited by slicing-based strategies in general, and that by \Vc
of multiplicative ones. 

For SHERPA, we used the COMIX~\cite{Gleisberg:2008fv} matrix-element
generator, while 
\Vc's matrix elements come from MADGRAPH~4~\cite{Alwall:2007st} and
HELAS~\cite{Murayama:1992gi}. A matching 
scale of 5 GeV was imposed for all matched multiplicities in
SHERPA. In \Vc, matching is normally carried out over all of
phase space; for this comparison, we 
limited the highest matched matrix elements to the region above 5 GeV, 
while lower multiplicities were still matched over
all of their respective phase spaces. In
both programs, bottom quarks 
were treated as massive, lighter ones as massless. The tests were run on
a single 3.05 GHz CPU (with 4GB memory) using gcc 4.6, with O2
optimization. Hadronization and initial-state photon radiation were
switched off.

The point about initialization time is clearly illustrated in the
left-hand pane of \figRef{fig:speed}; in the CKKW-based matching
strategy implemented in SHERPA, the integration
of each additional 
higher-leg matrix element and the warm-up of the 
corresponding phase-space generator takes progressively more time at
startup (note the 
logarithmic scale), while \Vc's initialization time is independent
of the desired matching level. 

In the right-hand pane of \figRef{fig:speed}, the time required to
generate (unweighted) events in CKKW also starts by rising rapidly,
but then eventually levels off and appears to saturate at $\sim$ 6 partons. We interpret the reason for this to be that, while
it still takes a long time to compute the total 6-jet cross section
(reflected in the left-hand pane), the actual value of that cross
section is quite small, and hence only a small fraction of the
generated events will actually be of the six-jet variety. The precise
behavior of course depends on the choice of matching scale. 

In addition, on the right-hand pane of \figRef{fig:speed}, 
the solid \Vc curves represent the new (helicity-dependent)
formalism, for global 
(triangle symbols) and sector (diamond symbols) showers,
respectively. The dashed curves shown in lighter shades give the corresponding
results without helicity dependence. At 2 partons, i.e., without any matching
corrections, we see that the \Vc showers are currently slightly
slower than the SHERPA ones. This was not the case in
\cite{LopezVillarejo:2011ap}, and is due to the trial-generation
machinery in \Vc having been rewritten in a simpler form,
which  is slightly more wasteful of random numbers, an optimization
point we intend to return to in the future. The main point of our
paper, however, is the scaling with the number of additional matched
legs exhibited by the helicity-dependent GKS matching formalism, which
is almost flat in the sector case, and still significantly milder in
the global case than for the CKKW-based SHERPA comparison.

\subsection{Validation}

To complete the validation of the new helicity-dependent framework, 
we include a set of comparisons to LEP measurements at 
the event-, jet-, and particle levels, respectively. These comparisons
were carried out using the default settings for \Vc 1.029, which
include a slight re-optimization of the hadronization parameters in a
new default tune called ``Jeppsson 5'', with parameters given in
\appRef{app:j5}. For reference, comparisons to default \Py 8.172 (with
\Vc switched off) are provided as well.
In all cases, we consider hadronic decays of unpolarized $Z$ bosons, 
at $E_{cm} = 91.2\,\mrm{GeV}$, corrected for initial-state photon
radiation effects, and letting particles with $c\tau > 100\,\mrm{mm}$
be stable. 

All plots were made using \Vc's ROOT-based runtime
displays~\cite{Giele:2011cb,Antcheva:2009zz}, which can be saved to
graphics files using the 
\texttt{VinciaRoot::saveDisplays()} command. \Vc is shown with solid
(blue) lines and filled dot symbols. \Py is shown with solid (red)
lines and open circle symbols. 
Experimental data is shown with black squares and black crosshairs
that correspond to one 
standard deviation. Where applicable, 
two crosshairs are overplotted on one another,
corresponding to statistical-only and total  
(stat + sys, summed in quadrature) uncertainties. 
Light-gray vertical extensions of the crosshairs illustrate two standard
deviations. The uncertainties on the MC runs are statistical 
only, and are shown at the two-sigma level, to be conservative. 
In the ratio panes below the main plots, we show theory/data; the
inner (green) shaded bands show one-sigma deviation contours, the
outer (yellow) ones two-sigma contours\footnote{For completeness, 
an additional very slight shading variation inside each band
shows the purely statistical component, where applicable.}.

\begin{figure}[t]
\centering
\includegraphics*[scale=0.25]{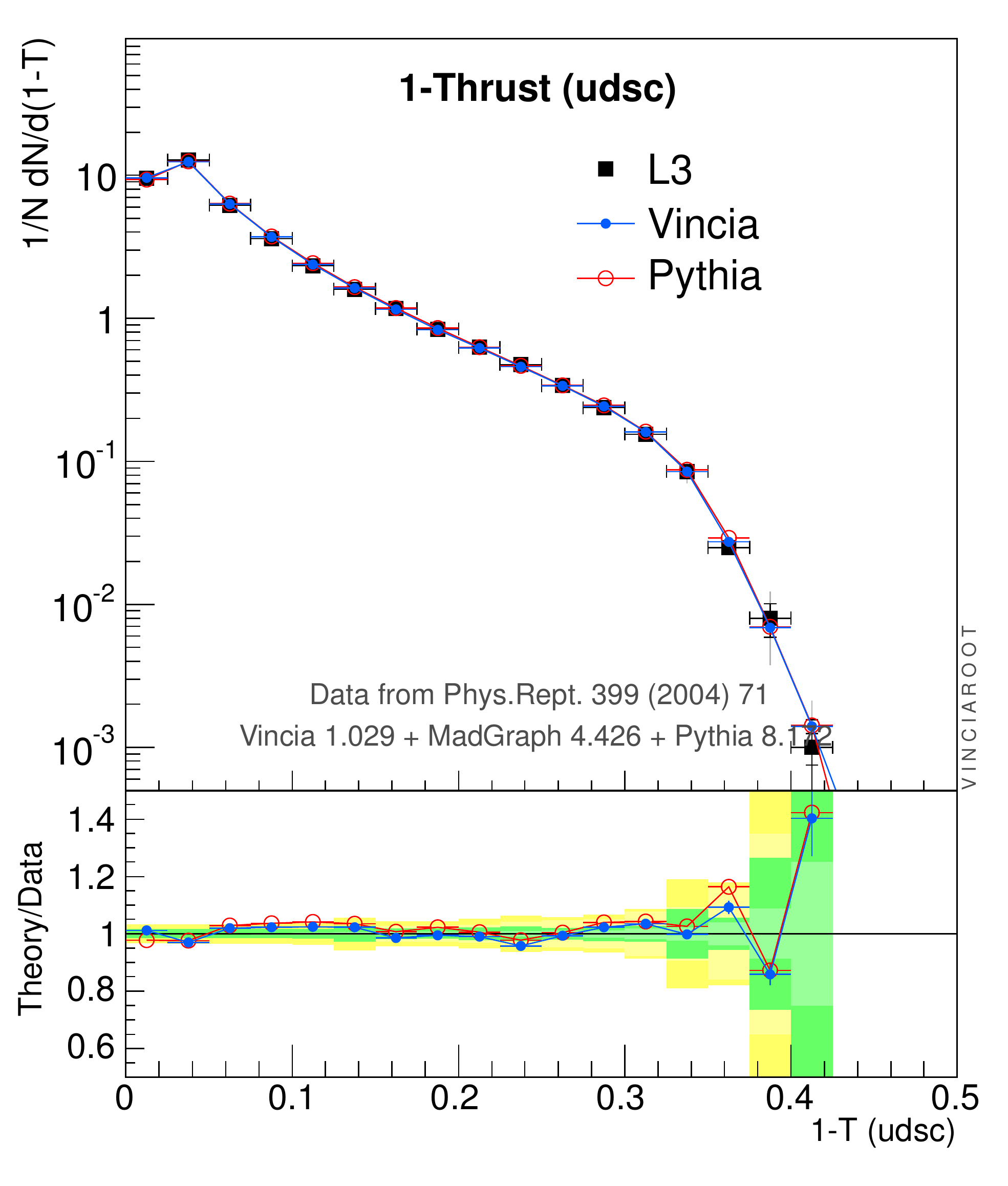}
\includegraphics*[scale=0.25]{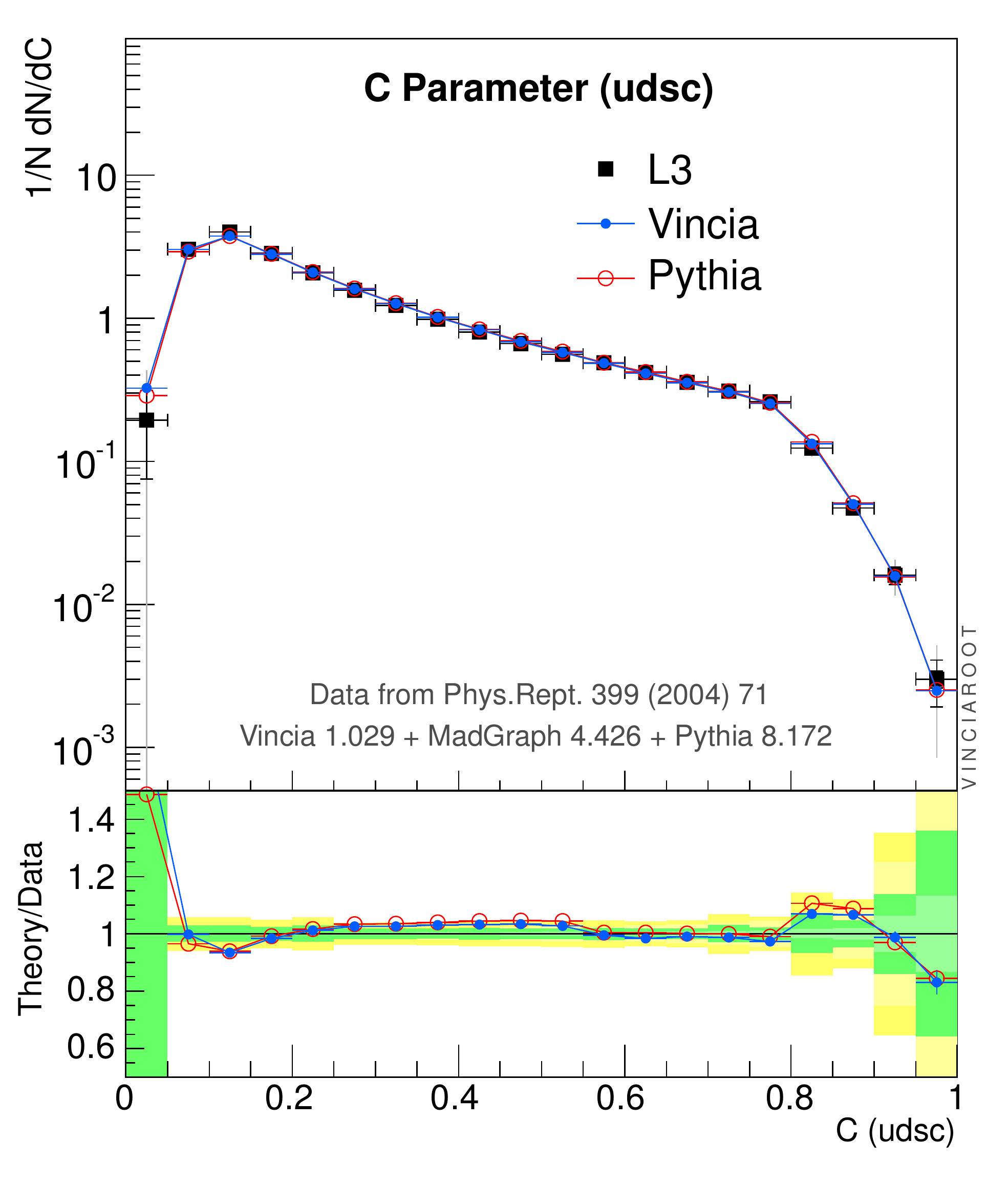}
\includegraphics*[scale=0.25]{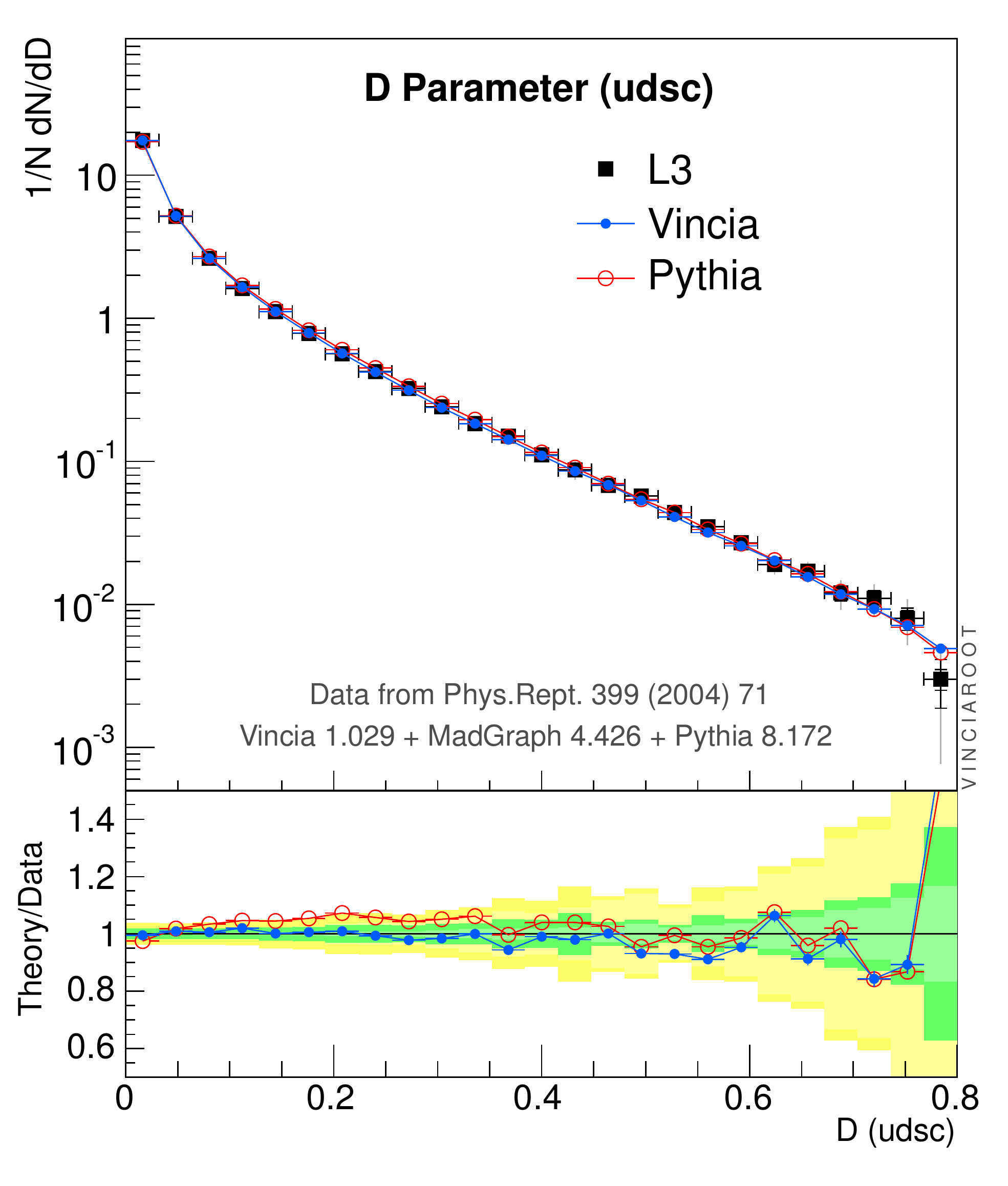}
\includegraphics*[scale=0.25]{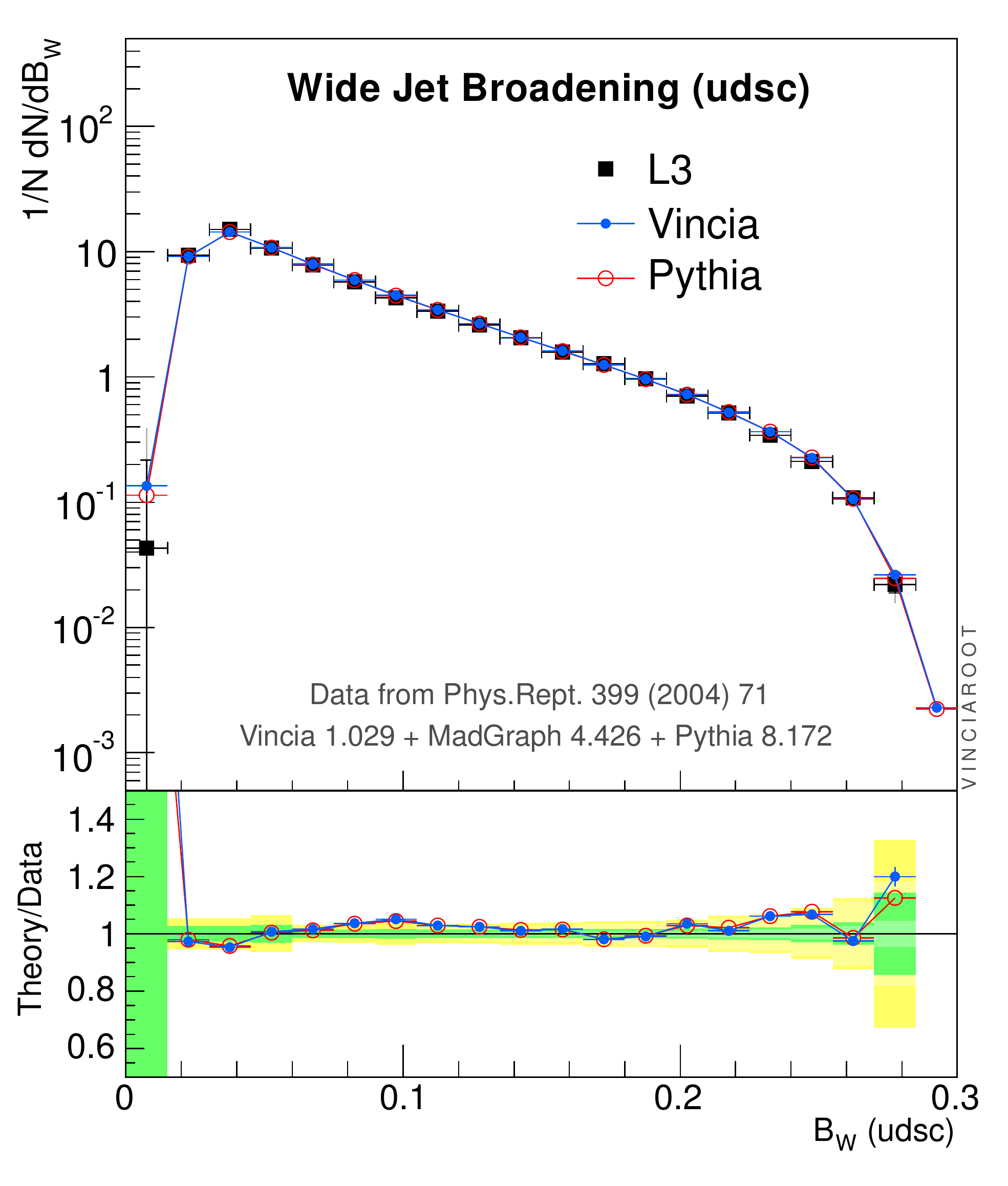}
\includegraphics*[scale=0.25]{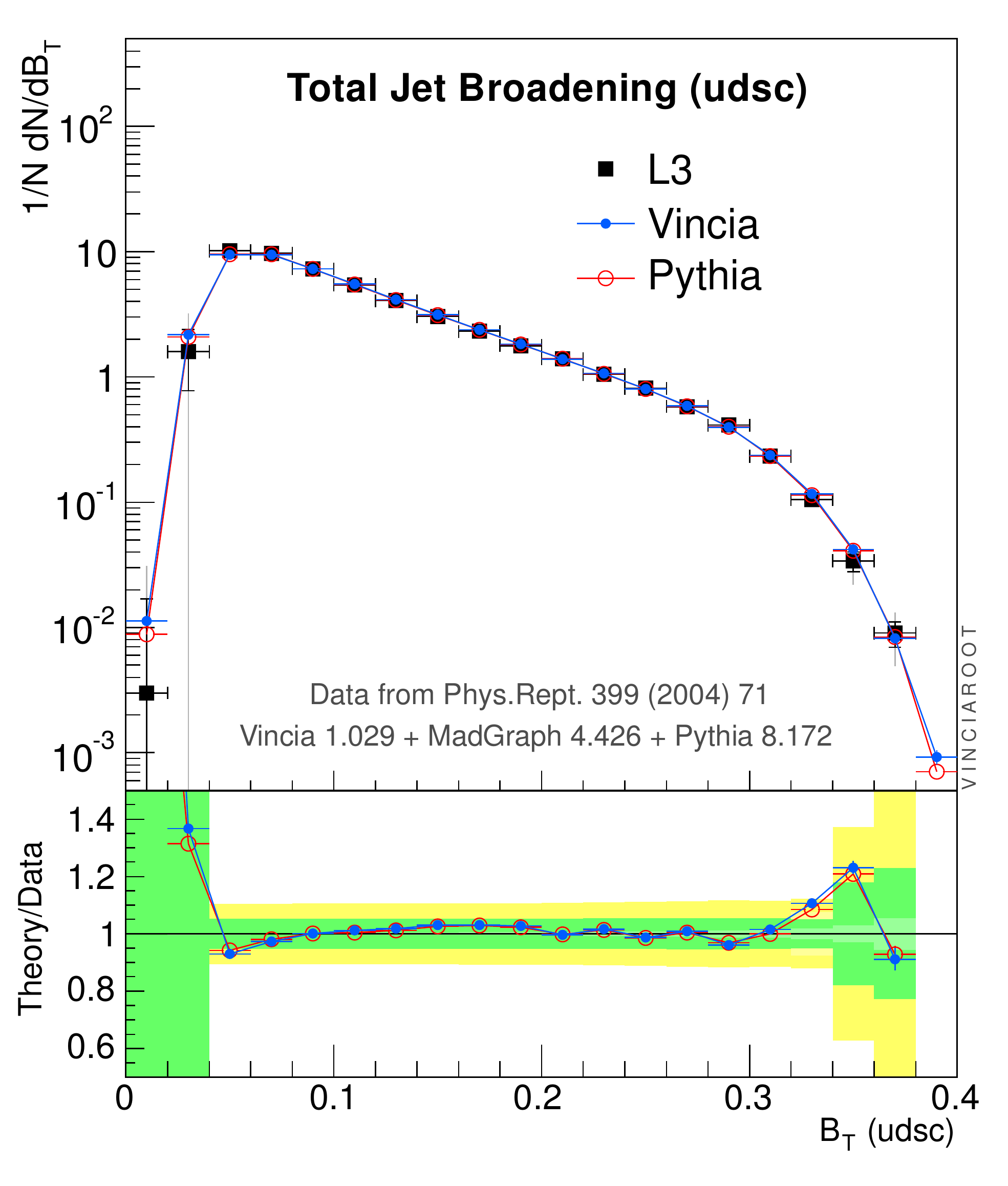}
\caption{Event shapes measured in light-flavor tagged events by the L3
  experiment at the $Z$ pole~\cite{Achard:2004sv}, 
  compared to default \Vc~1.029 and \Py~8.172. \label{fig:shapes}}
\end{figure}

Since we only apply the helicity-dependent formalism to massless
partons, we begin by focusing on light-flavor tagged
events. A very useful such set of measurements was performed by the L3
collaboration~\cite{Achard:2004sv}. In \figRef{fig:shapes}, we show
how default \Vc 1.029 compares to that data set, for the Thrust, C,
and D parameters (top row), and for the Wide and Total Jet Broadening
(bottom row), see ~\cite{Achard:2004sv} for definitions. No
significant deviations are observed, hence the code passes this
validation step. As in previous \Vc
studies~\cite{Giele:2011cb,LopezVillarejo:2011ap} (and \Py
ones~\cite{Buckley:2009bj,Skands:2010ak,Corke:2010yf}), 
however, one should note that this agreement comes at the price of
using a rather large value for $\alpha_s(M_Z)$,
\begin{equation}
\alpha_s(M_Z)~=~0.139~,
\end{equation}
which, with one-loop running (the default in \Vc), corresponds to a
five-flavor $\Lambda_\mrm{QCD}$ value of
\begin{equation}
\Lambda^{(5)}_{\mrm{QCD}} = 0.25~\mrm{GeV}~.
\end{equation}
In a pure parton shower, 
such a large value could perhaps be interpreted as an attempt to
compensate for missing hard higher-multiplicity matrix-element
corrections. With \Vc, however, we find that such an interpretation
cannot be the whole story, since the default \Vc settings include
LO matrix-element corrections through $Z\to 5$ partons. 

In our view,
there are two factors contributing to the large $\alpha_s$ value
favored by the \Py and \Vc tunings. Firstly, the $\alpha_s$
value extracted from a Monte Carlo tuning is not guaranteed to
be directly interpretable as an $\overline{\mrm{MS}}$ value. Indeed,
CMW argued~\cite{Catani:1990rr} that a rescaling of the effective
$\Lambda_{\mrm{QCD}}$ value by a factor 1.57 (for 5 flavors) 
is appropriate when translating from $\overline{\mrm{MS}}$ to a
coherent Monte-Carlo shower scheme. With the caveat that the original 
CMW argument was based on two-loop
running while \Vc currently defaults to one-loop running, 
a naive application to the value found above
would reduce the equivalent $\overline{\mrm{MS}}$ value to:
\begin{equation}
\Lambda^{(5)\overline{\mrm{MS}}}_{\mrm{QCD}} ~ \sim ~
\frac{0.25~\mrm{GeV}}{1.57}~=~ 0.16~\mrm{GeV}~,
\end{equation}
corresponding to a (one-loop) value for $\alpha_s(M_Z)$ of $\sim$
0.129. Secondly, 
this still rather high value should then be seen in the context
of an LO+LL extraction. The inclusion of the full NLO
correction to $Z\to 3$ jets, which is the topic of 
a forthcoming paper~\cite{ee3jets}, generates additional order 10\%
corrections to hard radiation, which should bring the extracted value
further down, and the expectation is that the tuned value would then
be in accordance with other NLO extractions. We shall return to this
issue in more detail in~\cite{ee3jets}.

\begin{figure}[t]
\centering
\includegraphics*[scale=0.25]{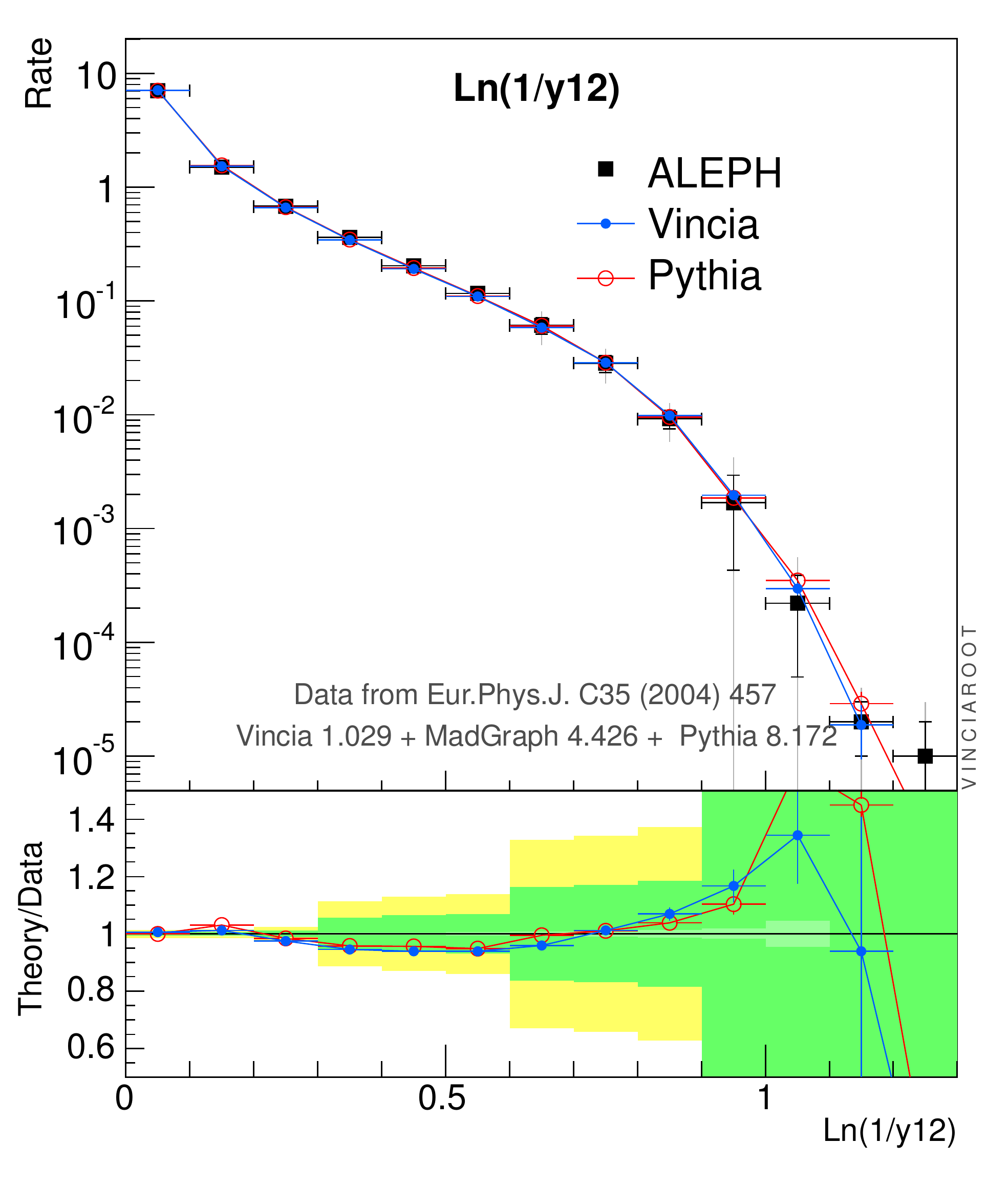}
\includegraphics*[scale=0.25]{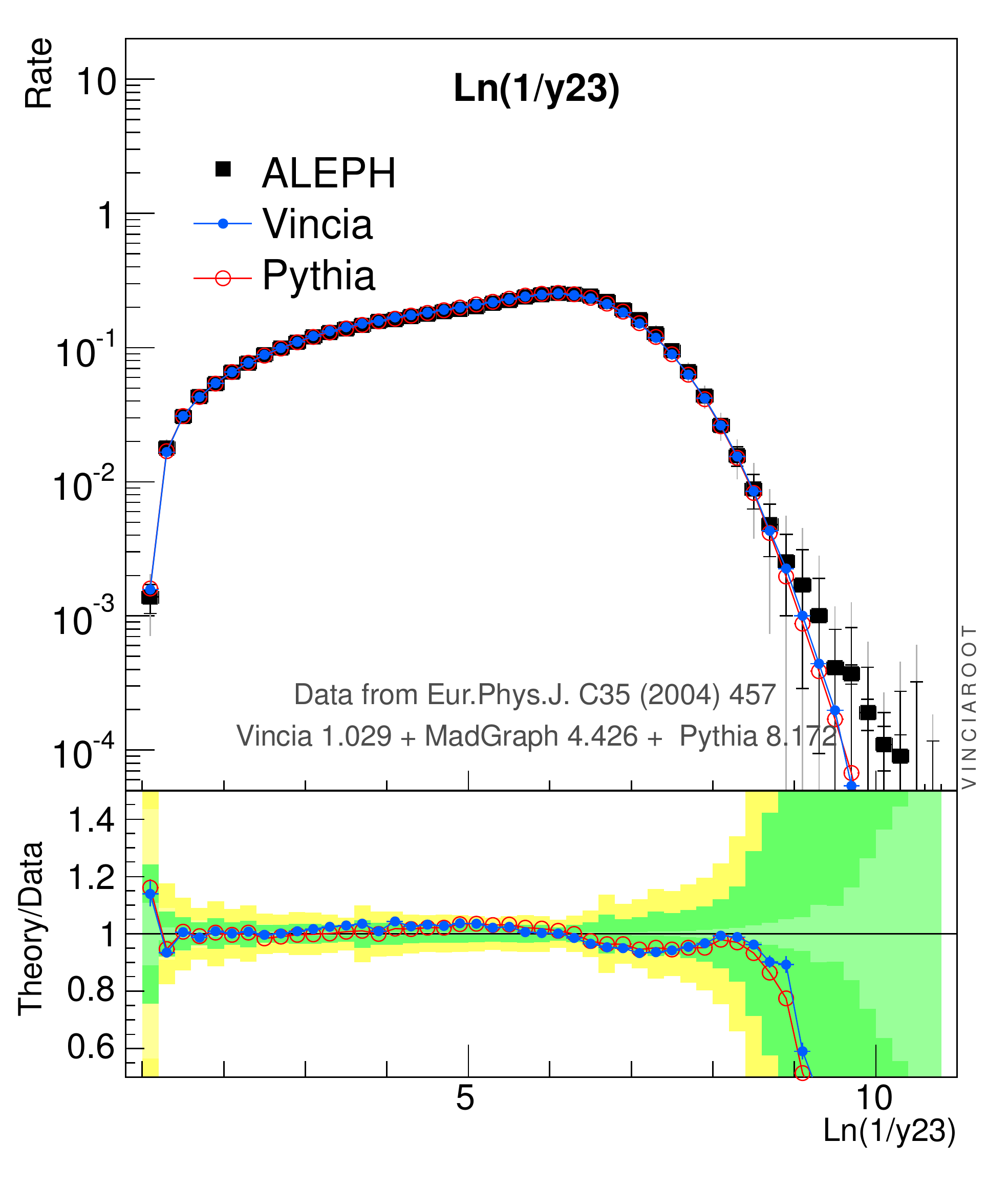}
\includegraphics*[scale=0.25]{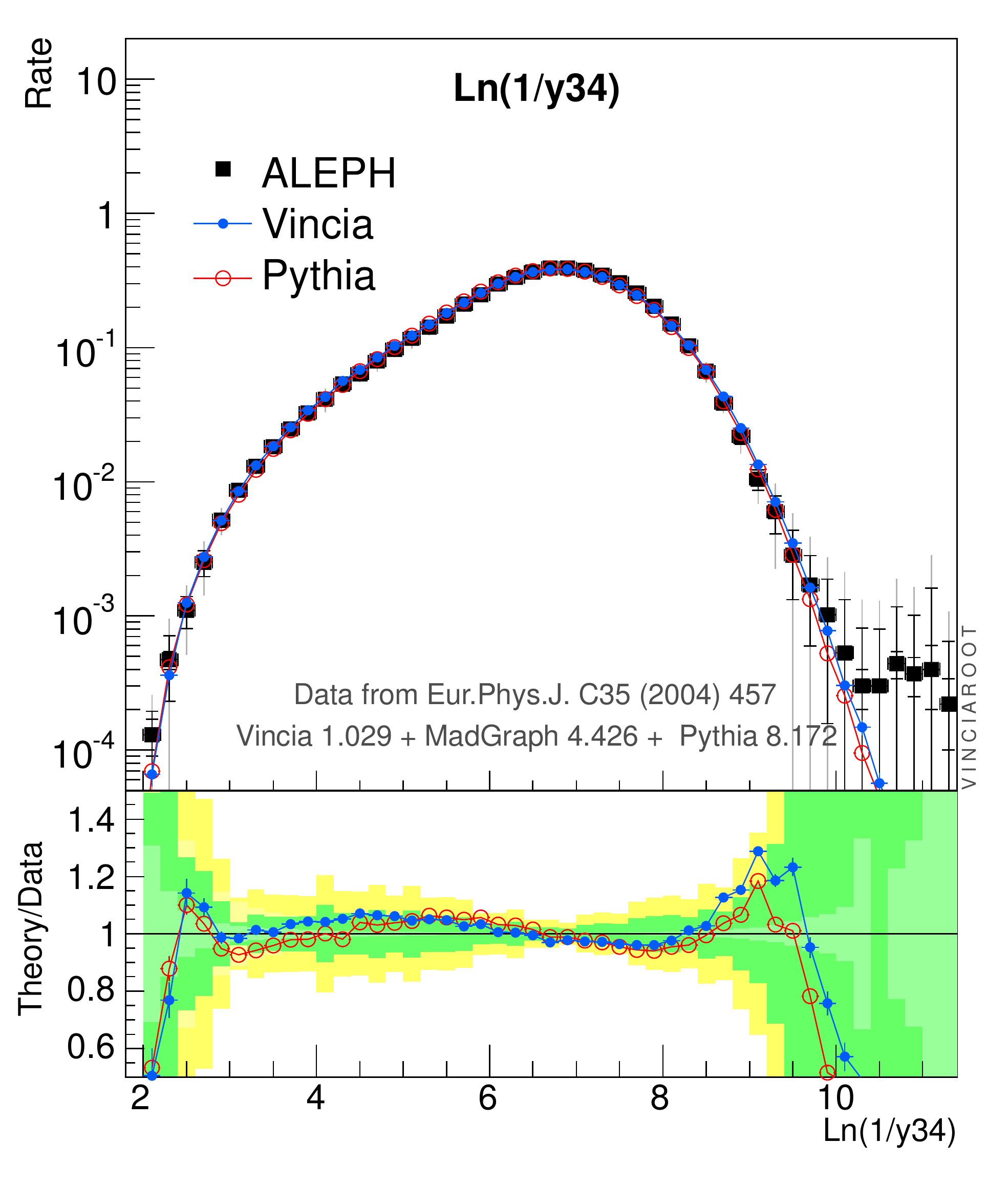}
\includegraphics*[scale=0.25]{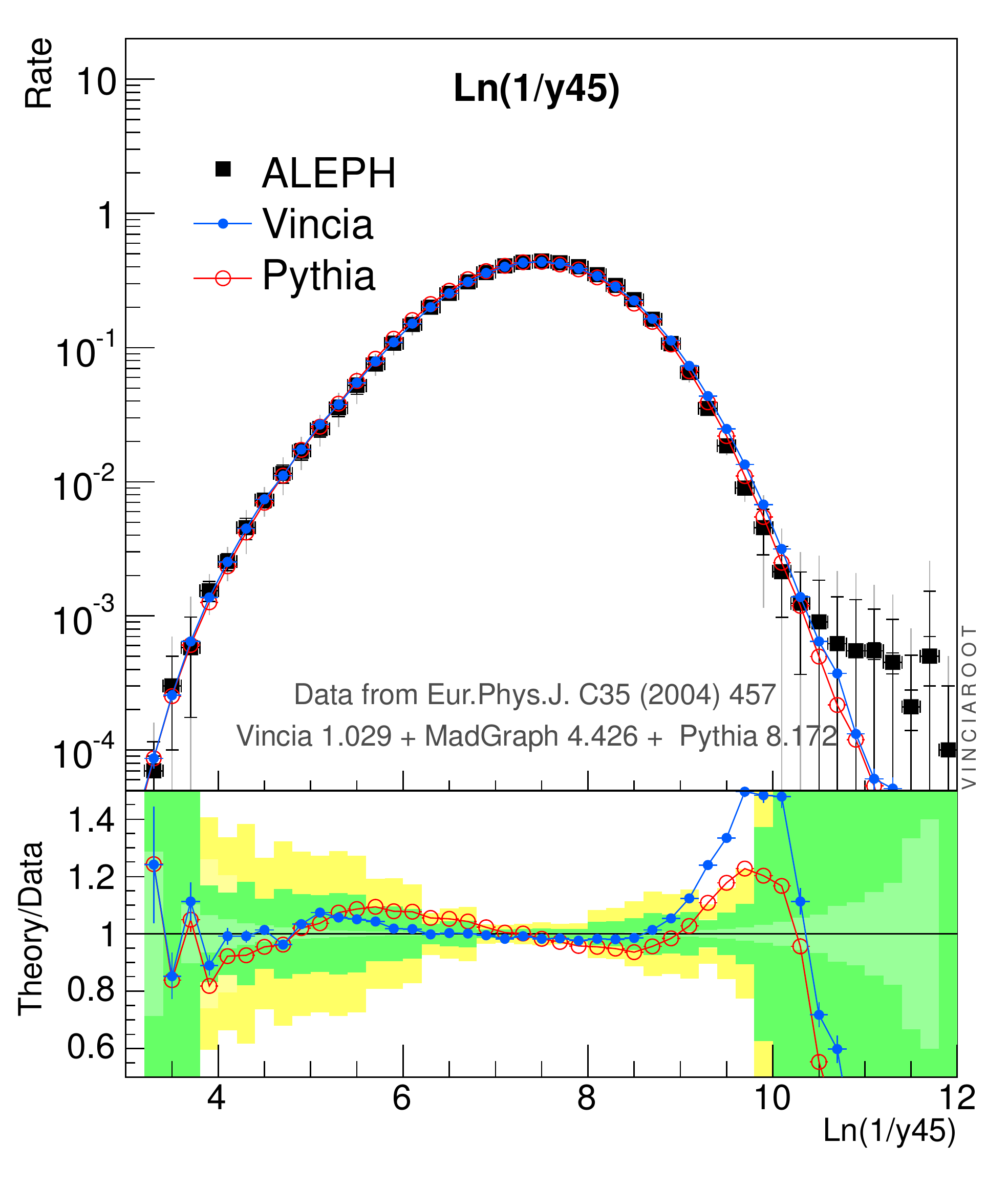}
\includegraphics*[scale=0.25]{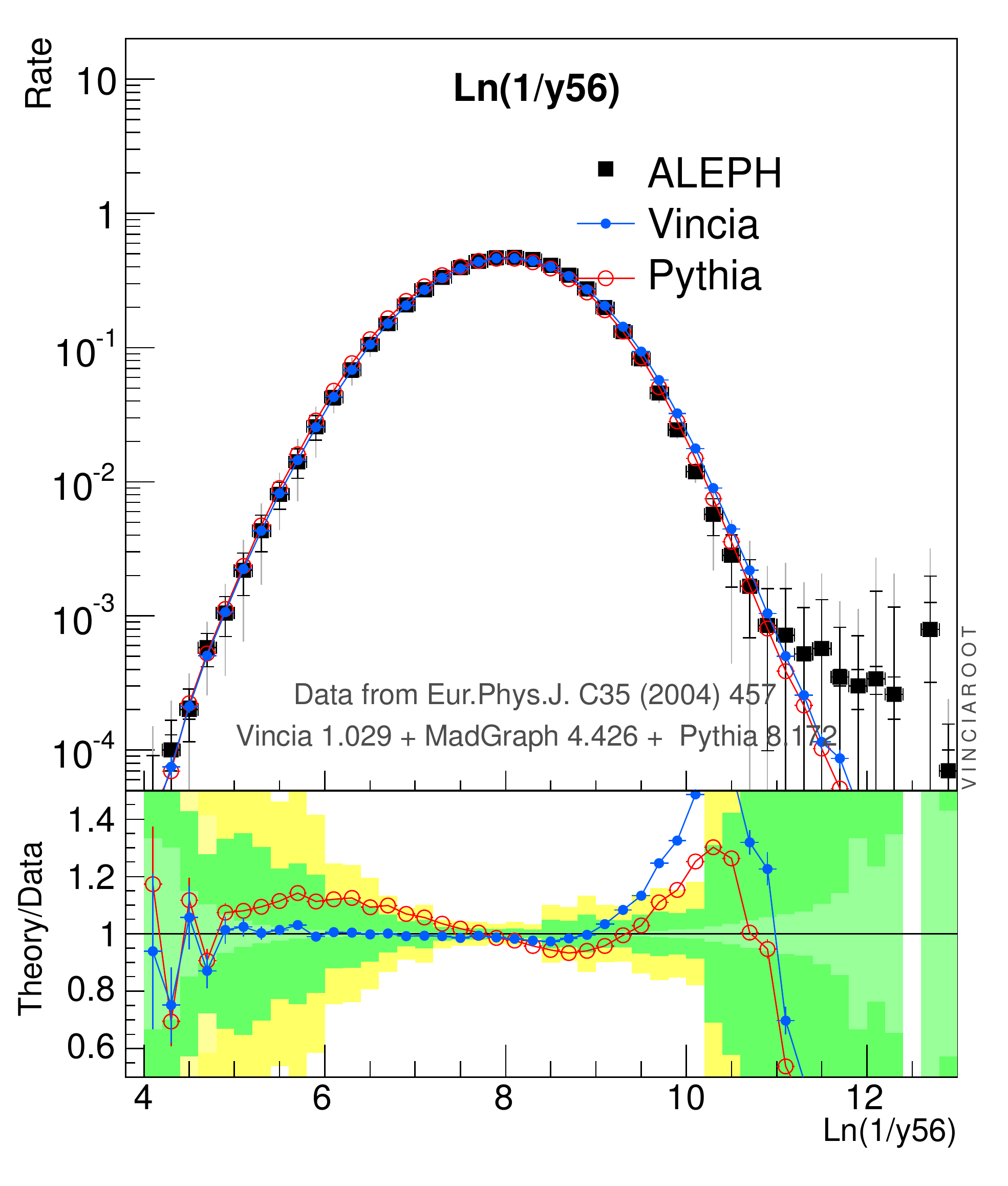}
\includegraphics*[scale=0.25]{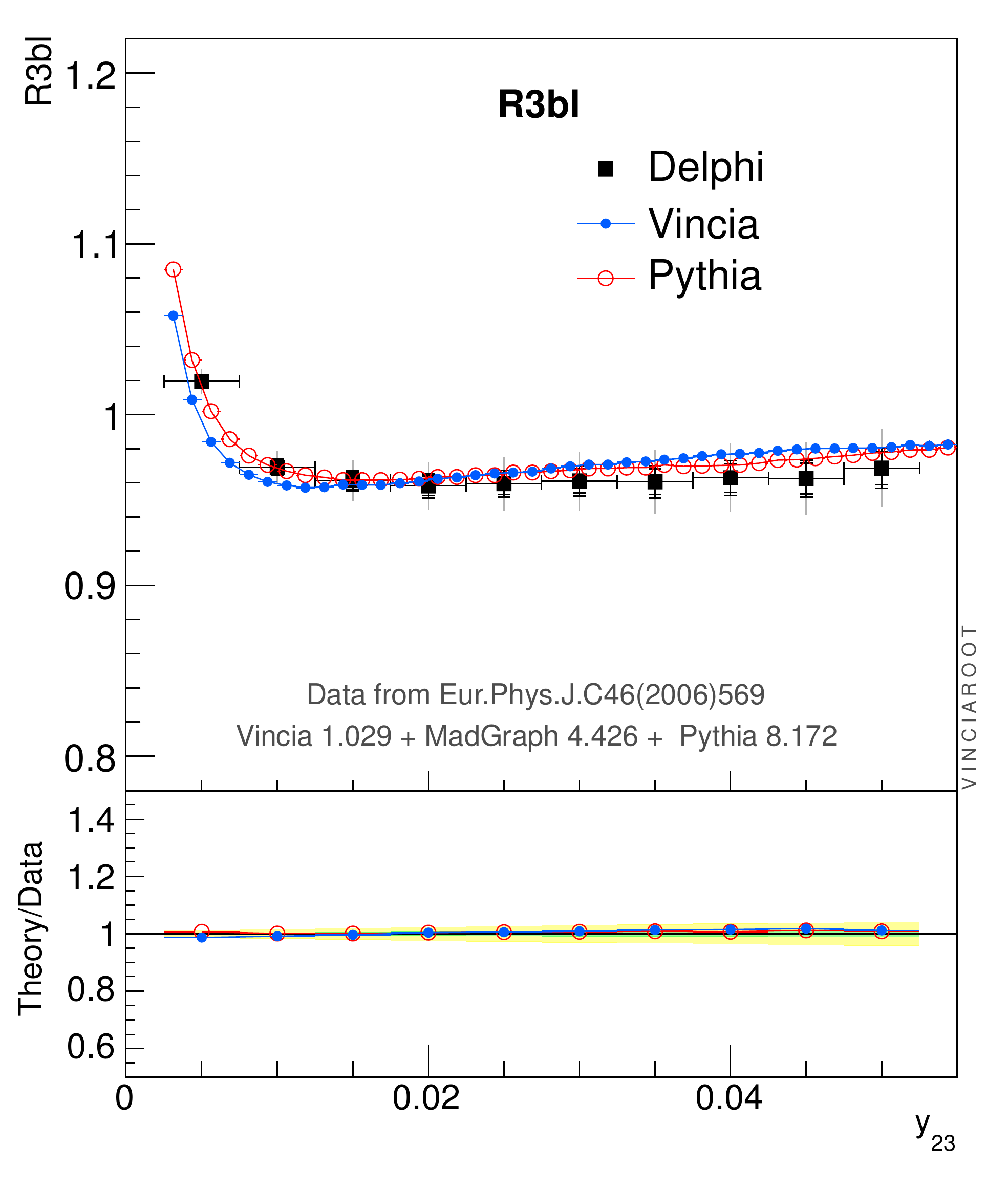}
\caption{{\sl From Top Left:} Jet resolution distributions 
  measured by the ALEPH experiment at the $Z$
  pole~\cite{Heister:2003aj}.
 {\sl Bottom Right:} The ratio of
  3-jet rates in $b$- vs.\ light-flavor tagged events, as a function
  of Durham $y_{23}$, measured by the
  DELPHI experiment~\cite{Abdallah:2005cv}.  
  Comparisons to default \Vc~1.029 and \Py~8.172. \label{fig:jets}}
\end{figure}

Passing now from event shapes to jets, the first 5 panes of 
\figRef{fig:jets} show a
comparison to the 2-, 3-, 4-, 5-, and 6-jet resolution scales measured
by the ALEPH collaboration~\cite{Heister:2003aj} (now including also
$Z\to b\bar{b}$ events), using the Durham $k_T$ clustering
algorithm~\cite{Catani:1991hj}, with distance measure
\begin{equation}
y_{ij} \ = \ \frac{2\mrm{min}(E_i^2,E_j^2)(1-\cos\theta_{ij})}{E_\mrm{vis}^2}~,
\end{equation}
for which we use the FASTJET
implementation~\cite{Cacciari:2011ma}. Formally, $E_\mrm{vis}$ is the
total visible energy, but since the ALEPH data were corrected for the
distortions caused by neutrinos escaping detection, we here include
neutrinos in the inputs passed to FASTJET. Hard scales have values 
$\ln(y)\sim 0$ and hence appear towards the left edge of the plots in
\figRef{fig:jets}, while soft scales appear towards the right-hand
edges. Non-perturbative effects are expected to dominate below roughly
1 GeV, corresponding to $\ln(1/y)\sim\ln(91^2/1^2) \sim 9$. Above this
scale, i.e.\ in the perturbative region, 
we observe no disagreement between the ALEPH data and \Vc. 
(Note that the distributions are affected by the kinematics of $B$
decays starting already from $\ln(1/y)\sim\ln(91^2/5^2) \sim 5.8$, but
these decays are modeled adequately by \Py, and hence do not
trouble this comparison. The feature at $\ln(1/y)\sim 10$ in the 5-
and 6-jet resolutions corresponds to clustering scales below 1 
GeV and hence is likely to be associated with a combination of string
breaking and hadron decays.)

As a verification that the perturbative mass corrections for heavy
quarks have not been altered by the new implementation, 
the last pane in
\figRef{fig:jets} shows the ratio of $b$- to light-quark 3-jet
resolutions measured by DELPHI~\cite{Abdallah:2005cv}, 
which also appeared as one of the validation plots in our dedicated 
study of mass effects~\cite{GehrmannDeRidder:2011dm}. The distribution
is essentially unchanged with respect to the previous study, and
retains its non-trivial shape. 

\begin{figure}[t]
\centering
\includegraphics*[scale=0.25]{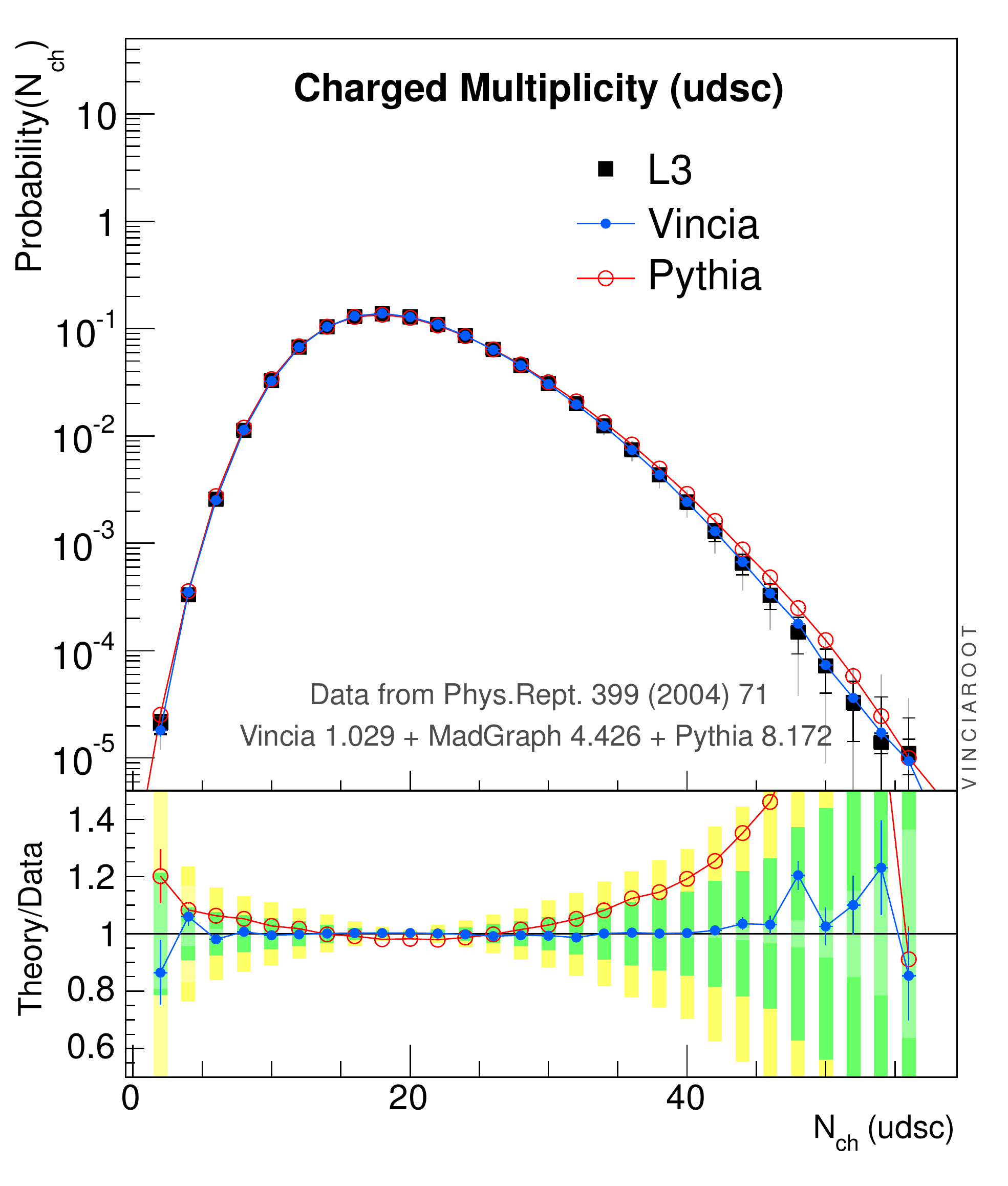}
\includegraphics*[scale=0.25]{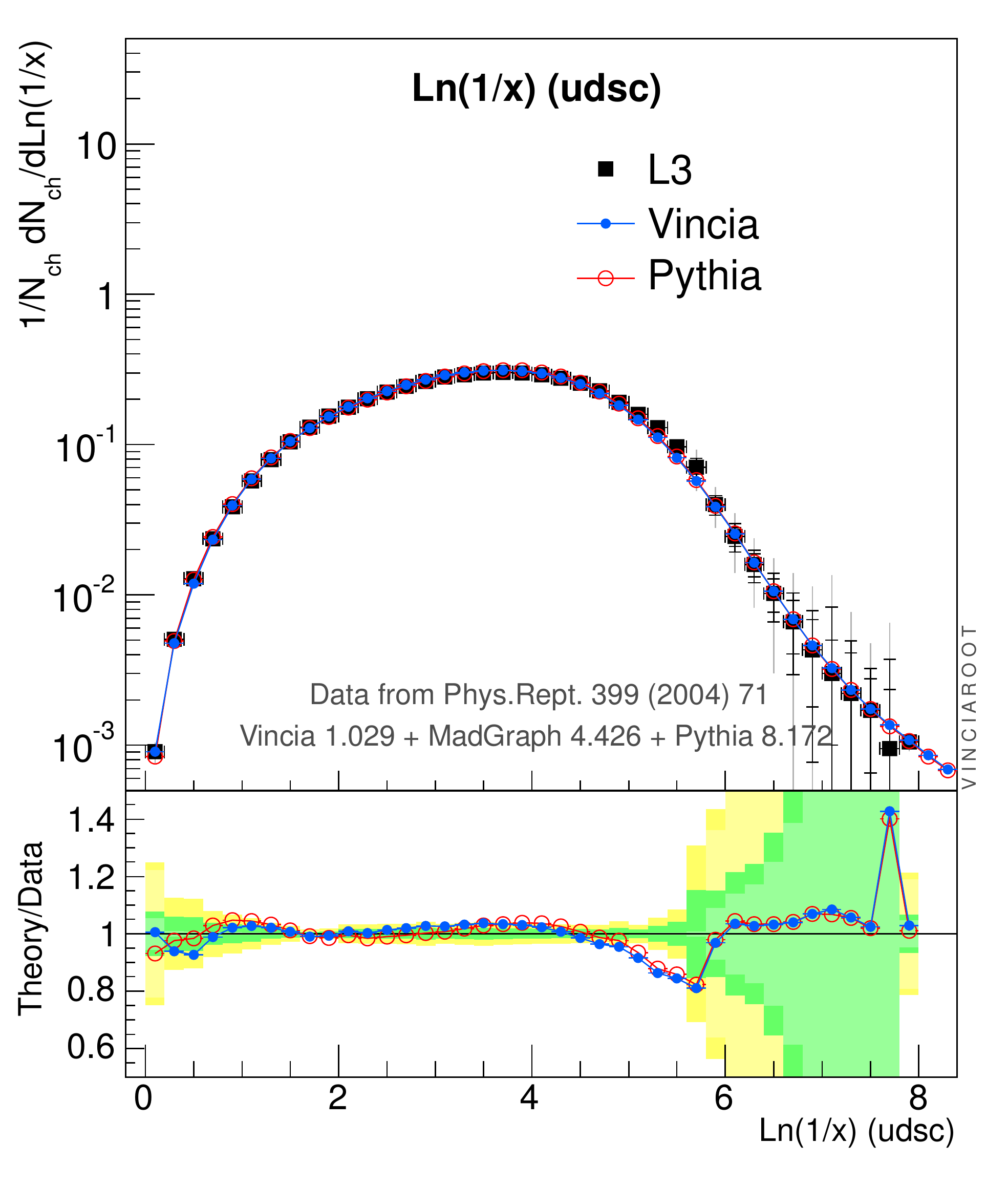}
\\
\includegraphics*[scale=0.25]{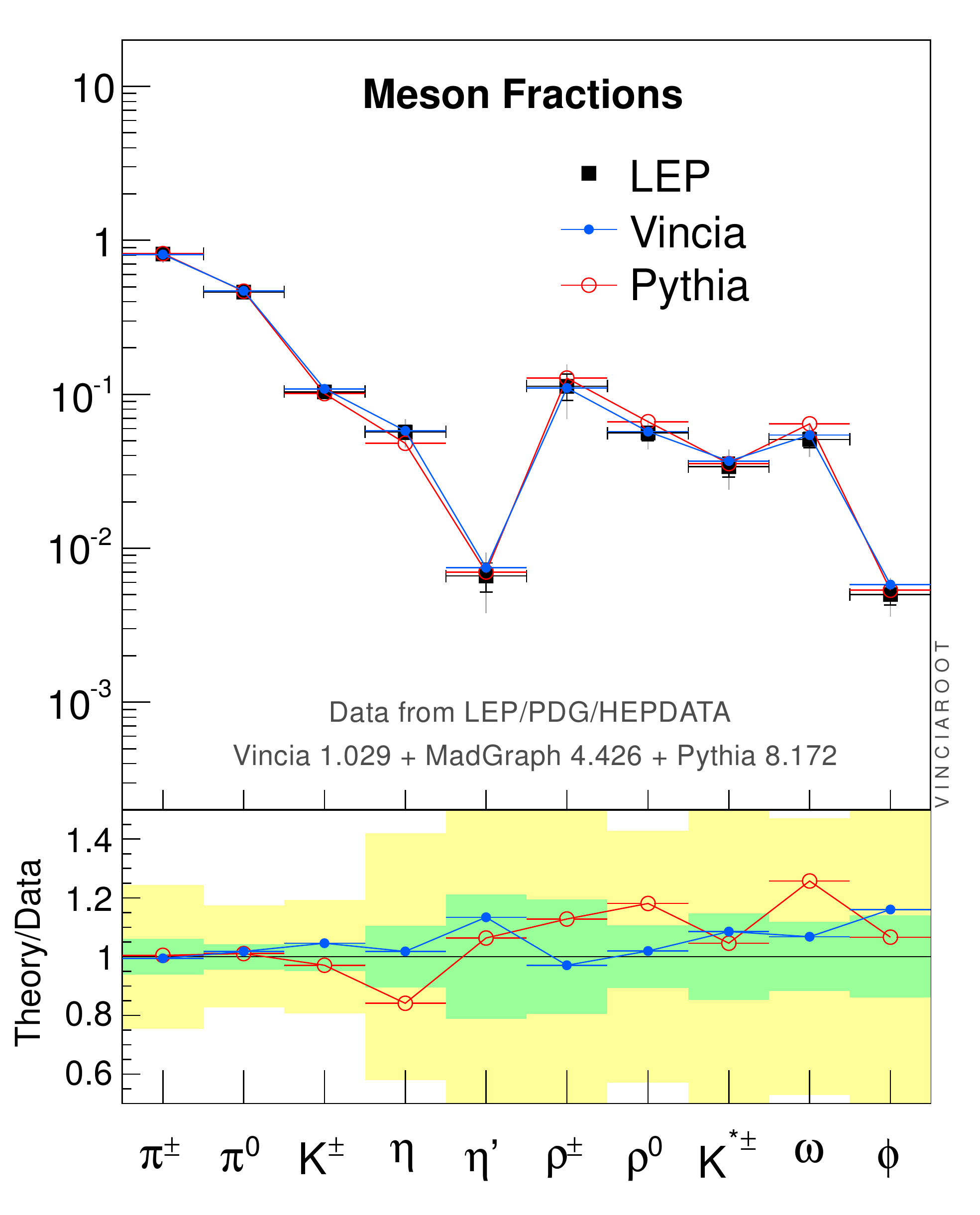}
\includegraphics*[scale=0.25]{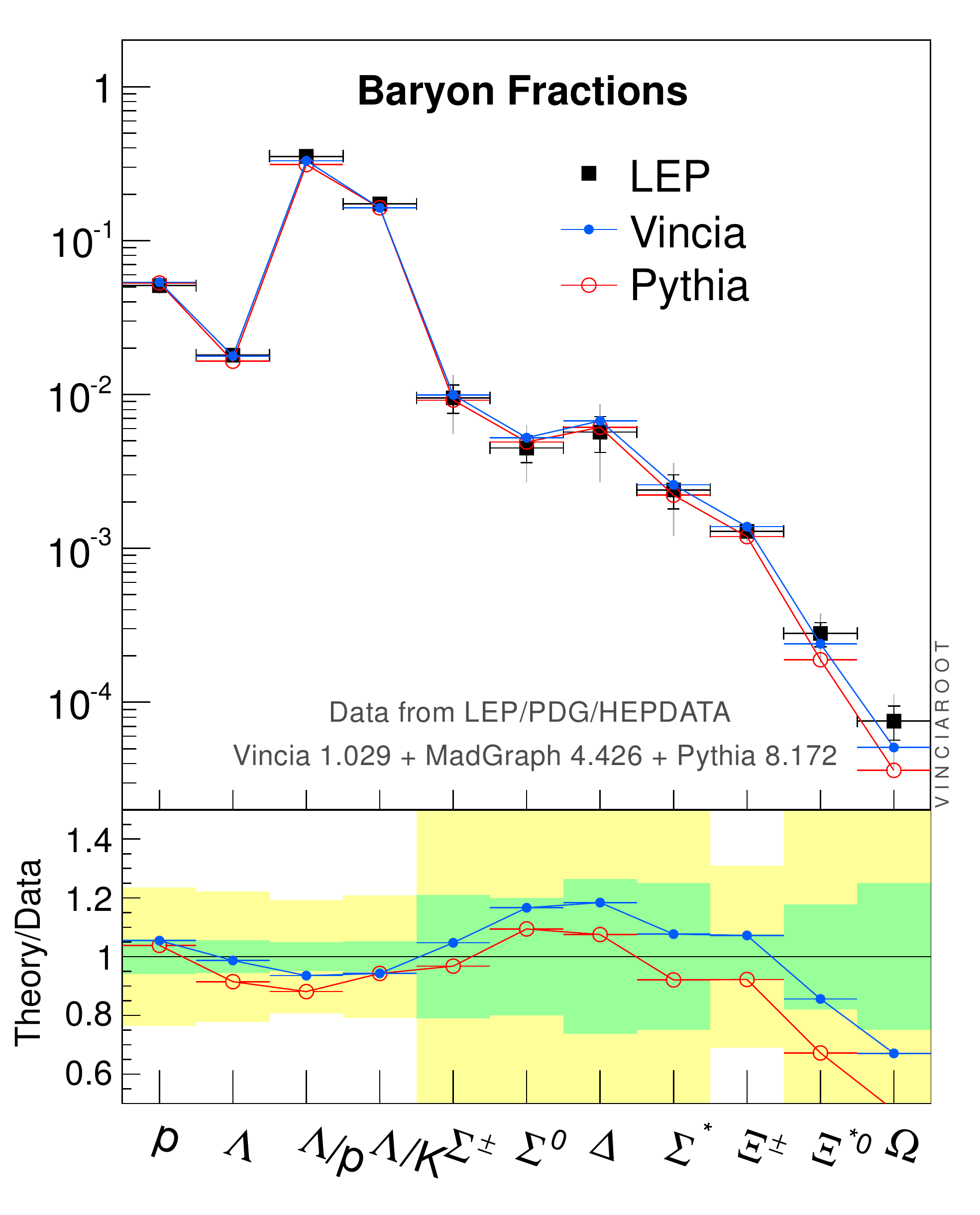}
\includegraphics*[scale=0.25]{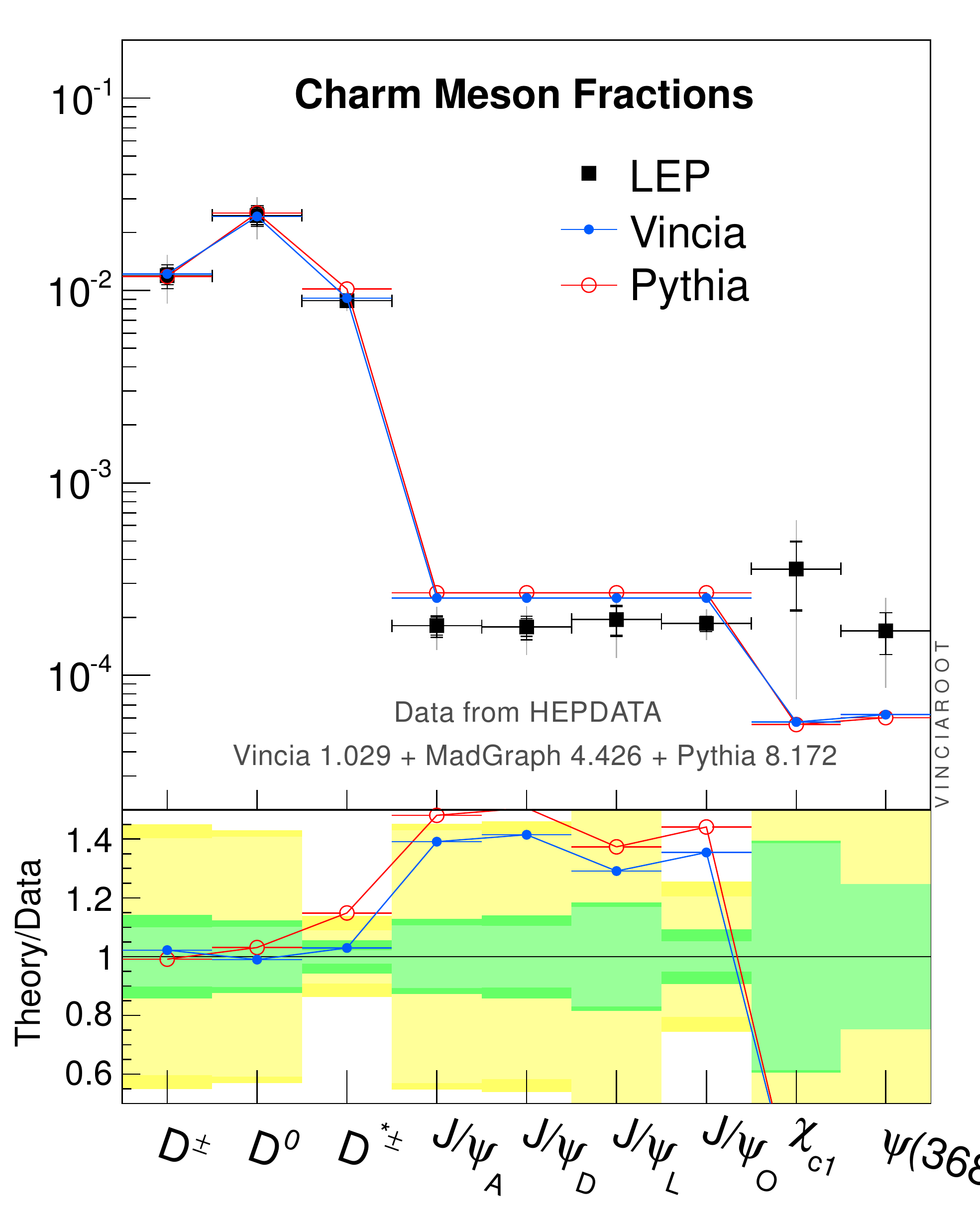}
\caption{{\sl Top Row:} Inclusive charged-particle multiplicity and
  momentum spectra in 
  light-flavor tagged events measured by the L3 experiment at the $Z$
  pole~\cite{Achard:2004sv}. 
  {\sl Bottom Row:} Meson, Baryon, and Charm-Meson fractions
  (normalized to the average charged multiplicity. Comparisons to
  default \Vc~1.029 and \Py~8.172. \label{fig:particles}}  
\end{figure}

Lastly, we turn to distributions at the individual particle
level. The top row of \figRef{fig:particles} shows the 
charged-particle multiplicity and momentum spectra, again for
light-flavor tagged L3 events~\cite{Achard:2004sv}, with no significant
deviations between \Vc and the data. (The feature around
$\ln(1/x)\sim 6$ corresponds to momentum scales close to the pion mass
and is also seen in standalone \Py, hence we interpret it as an issue
with the non-perturbative hadronization model.) 

The bottom row of \figRef{fig:particles} shows the relative fractions
of various identified particles, normalized by the average
charged-particle multiplicity. The experimental numbers are here
labeled ``LEP'' and represent our own estimates, using a combination
of inputs from PDG~\cite{Beringer:1900zz} and
HEPDATA~\cite{Buckley:2010jn}. The two leftmost panes show meson and
baryon fractions, respectively. The meson fractions are somewhat better
described than the baryon ones, and slightly different tuning
priorities are evident between \Py and \Vc,
but in no case do we see a significant
deviation from the data. One remark is worth making, though, that the
production of strange and multi-strange baryons tends to be at the
lower limit of what is allowed by the data. We have addressed this by
removing any strange-baryon suppression relative to light-flavor ones 
in the string-fragmentation flavor selection,
but note that the data might even prefer a slight enhancement,
which is currently not a technical possibility in \Py.

As an additional piece of information less
relevant to the study performed here, the last pane of
\figRef{fig:particles}  shows a comparison
to charm meson fractions. Though the total amount of charm meson production is
reasonably well described, there appears to be a slight overproduction of
$D^*$ mesons in \Py, and both \Vc and \Py exhibit an excess of
$J/\psi$ resonances (the four data points correspond to each of the
four LEP experiments) and an 
underproduction of $\chi_{c1}$ and $\psi(3685)$. Especially the latter
rare states are
certainly not expected to be perfectly described out of the box, and hence
we mostly include this comparison as a hint of where future
improvements might be useful. 

Finally, we should also mention that the code performs several
internal self-consistency checks during initialization. In particular,
the soft and collinear limits of all antenna functions are checked
against the respective eikonal and (helicity-dependent) 
Altarelli-Parisi kernels, and a
verification is made that the antenna functions remain positive over
all of the physical phase space.

\section{Conclusions \label{sec:conclusion}}

Our development of a helicity-based shower in \Vc shows that significant speed gains are obtained when matching to helicity matrix elements
as compared to matching an unpolarized shower to spin-summed matrix elements.  One reason for this is that the MHV-type
helicity configurations tend to be the dominant contribution to the spin-summed matrix element.  MHV amplitudes are also those
best described by the shower because they contain the maximum number of soft and collinear singularities.  In addition, the intrinsic 
accuracy of the helicity shower is increased with respect to the unpolarized shower for essentially the same reason.  

There are several directions in which the helicity formalism developed here can be extended in \Vc.  First, as mentioned
in~\secRef{sec:mass}, mass effects can be included using the phase-space maps from~\cite{GehrmannDeRidder:2011dm} 
and the massive splitting functions from~\cite{Larkoski:2011fd}.  In the massive case, the spin of a particle does not
have an unambiguous definition and so one must take care in defining the spin in a consistent manner.  We advocate for
defining the spin of a massive fermion by its chirality; however, the spin can also be defined by a projection onto a reference
vector.  Both have subtleties: chirality is Lorentz invariant, but can flip from mass insertions.  Using a reference vector to define
spin breaks Lorentz invariance so one must be careful to use the same reference vector for all calculations so that the final
result, when summed over spins, is Lorentz invariant.  Our advocation for using chirality is based on its Lorentz invariance
as well as its importance in weak decays.  Mass effects are particularly important in top quark decays where all of these
effects can be studied.

While there are many subtleties in extending \Vc to include initial state radiation (ISR) for hadron collisions~\cite{Ritzmann:2012ca} or next-to-leading
order (NLO) matching~\cite{ee3jets}, 
implementing the helicity shower within these
frameworks should be straightforward.  The helicity antennae 
described here would work in the initial state as well, with the possible caveat that the finite terms in the antennae might need to be changed
to guarantee positivity in the initial state phase space.  Matching the shower in hadron collisions to helicity amplitudes would maintain
the speed gains illustrated here.  However, the matching procedure would have to be changed because, for hadron collisions, it would no
longer be practical to develop a complete library of matrix elements to which to match.  A hybrid approach in which matrix elements are
computed dynamically as well as extracting some matrix elements from libraries would be necessary.  A similar procedure exists in \Sh,
where tree-level helicity amplitudes are computed from Berends-Giele~\cite{Berends:1987me} recursion relations in COMIX~\cite{Gleisberg:2008fv} 
and by Feynman diagrams in AMEGIC++~\cite{Krauss:2001iv}.  Matching to NLO matrix elements would require utilizing the libraries 
from BLACKHAT~\cite{Berger:2008sj}, for example, in which the individual helicity components to the process can be extracted.  A recent 
study~\cite{Badger:2012uz} showed that the most efficient method for calculating helicity amplitudes depends on the spin configuration as well as the
number of external particles. 

\subsection*{Acknowledgments}

This work was supported in part by the European Commission (HPRN-CT- 200-00148), 
FPA2009-09017 (DGI del MCyT, Spain) and S2009ESP-1473 (CA Madrid). J.J. L-V has been 
supported by a MEC grant, AP2007-00385 and as a CERN visitor at the Theory Division. He wants to thank its hospitality, and also the groups at CEA Saclay IPhT and Lund.  
A.~L.~is supported by the US Department of Energy 
under cooperative research agreement DE--FG02--05ER41360 and supported in part by the U.S. National 
Science Foundation, grant NSF-PHY-0969510, the LHC Theory Initiative, Jonathan 
Bagger, PI.  A.~L.~also wishes to thank the CERN Theory Division and the CEA 
Saclay IPhT for their hospitality during recent visits.
 
 \appendix

\section{Antenna Construction}

\subsection{\label{app:globalunpol}Construction of Unpolarized Global Antennae}

In this appendix, we present the construction of the unpolarized global antennae
as an illustration of the more general procedure for determining the
helicity-dependent global antennae.
To determine the singular terms in spin-dependent global antenna functions, 
it suffices to consider the $gg\to ggg$ emission and $gg\to g\bar{q}q$ splittings only.
The gluon splitting antennae are the same as their sector
counterparts, decreased by a 
factor of 2 because of the two antennae which contribute.  The global
gluon emission antennae 
are significantly more complicated.
 We first use the constraints described in \secRef{subsec:GlobalAntennae} to
determine the spin-summed global antenna functions to see how they can be used to
reproduce familiar results.  The procedure will naturally generalize to the spin-dependent case.

To analyze the collinear limits of two gluons $j$ and $k$ it suffices to consider the configuration of four ordered gluons
 $i$, $j$, $k$ and $l$.  There are two splittings that contribute to the collinear limit of $j$ and $k$:
\begin{align*}
i \  (\hat{j}\ \hat{l}) &\to i \ (j \ k \ l) \ ,\\
(\hat{i} \  	\hat{k})\ l &\to (i \ j \ k) \ l \ .
\end{align*}
Here, the parentheses associate the $2\to 3$ splitting in the gluon configuration.  The singular terms of the
antenna for the first splitting can be written as
\begin{equation}
\hat{j}\ \hat{l} \to j \ k \ l = \frac{2}{y_{jk}y_{kl}}+\frac{f_1(y_{kl})}{y_{jk}}+\frac{f_2(y_{jk})}{y_{kl}},
\end{equation}
for some polynomials $f_1$ and $f_2$.  Similarly, the second splitting can be expressed as
\begin{equation}
\hat{i}\ \hat{k} \to i\ j \ k  = \frac{2}{y_{ij}y_{jk}}+\frac{f_3(y_{jk})}{y_{ij}}+\frac{f_4(y_{ij})}{y_{jk}} \ ,
\end{equation}
for some polynomials $f_3$ and $f_4$.  Note that by Bose symmetry, $f_1 = f_3$ and  $f_2 = f_4$.
Further, all $f_i$s are actually equal because of the symmetric initial and final splitting states.  
Thus, we will replace $f_i \equiv f$.

Now, consider the limit where $j\parallel k$.  In this limit, we set $y_{ij} = z$ and $y_{kl} = 1-z$.  
Then, the two antennae can be written as
\begin{eqnarray}
\hat{j}\ \hat{l} \to j \ k \ l &=& \frac{1}{y_{jk}}\left[ \frac{2}{1-z} + f(1-z)\right] \ , \\
\hat{i}\ \hat{k} \to i\ j \ k  &=& \frac{1}{y_{jk}}\left[ \frac{2}{z} + f(z)\right] \ .
\end{eqnarray}
Without loss of generality, we can write $f(z)$ in the form
\begin{equation}
f(z) = (-2-\alpha) + \alpha_1 z +\alpha_2 z^2 \ ,
\end{equation}
for some coefficients $\alpha$, $\alpha_1$ and $\alpha_2$.  For consistency, the sum of these 
splitting amplitudes must reproduce the Altarelli-Parisi splitting function:
\begin{equation}
P_{gg\leftarrow g}(z) = 2\left[ \frac{1-z}{z} + \frac{z}{1-z} + z(1-z)\right] = \left[ \frac{2}{1-z} + f(1-z)\right] + \left[ \frac{2}{z} + f(z)\right] \ .
\end{equation}
This requirement enforces $\alpha_2 = -1$ and $\alpha_1 = 1+2\alpha$ so that $f(z) = (-2-\alpha) + (1+2\alpha)z-z^2$.
Note that the GGG partitioning of global antenna \cite{GehrmannDeRidder:2005cm} corresponds to $\alpha = 0$ while 
the \Ar partioning \cite{Gustafson:1987rq,Lonnblad:1992tz} corresponds to $\alpha = 1$.  However, both are special cases of a one-parameter family of possibilities.

Positivity of the splitting function in the singular regions can be studied by taking, for example, the
limit of the antenna $\hat{i}\hat{k}\to ijk$ where $y_{jk}\to 0$ and $y_{ij}\to 1$.  In this limit, for the 
antenna to be non-negative, the function $f$ must satisfy $2+f(1)\geq0$ or that $\alpha\geq 0$.
Then, the global antenna for gluon emission is
\begin{equation}
a(gg\to ggg)  = \frac{2}{y_{ij}y_{jk}}+\frac{-2-\alpha + (1+2\alpha)y_{jk}-y_{jk}^2}{y_{ij}}+\frac{-2-\alpha + (1+2\alpha)y_{ij}-y_{ij}^2}{y_{jk}} \ .
\end{equation}
For $0\leq \alpha\lesssim4$, this antenna is positive on all of final-state phase space without the addition of any non-singular terms.
A table of the Laurent coefficients for generic partitioning of the unpolarized 
global antennae is presented in \tabRef{tab:summed_antennae}.

\begin{table}[htdp]
\begin{center}
\begin{tabular}{c c | c c c c c c }
$\times$ & $\frac{1}{y_{ij}y_{jk}}$ & $\frac{1}{y_{ij}}$ & $\frac{1}{y_{jk}}$ & $\frac{y_{jk}}{y_{ij}}$ & $\frac{y_{ij}}{y_{jk}}$ & $\frac{y_{jk}^2}{y_{ij}}$ & $\frac{y_{ij}^2}{y_{jk}}$   \\
\vspace{-0.4cm} \\
\hline
$q \bar{q} \to q g \bar{q}$ & $2$ & $-2$ & $-2$ & $1$ & $1$ & 0 & 0   \\
$q g \to qgg$  & $2$ & $-2$ & $-2-\alpha$ & $1$ & $1+2\alpha$ & 0 & $-1$ \\
$g g \to ggg $ & 2 &  $-2-\alpha$ &  $-2-\alpha$ & $1+2\alpha$ & $1+2\alpha$ & $-1$ & $-1$      \\
$q g \to q \bar{q} q$   & 0 & 0 & $\frac{1}{2}$ & 0 & -1 & 0 & $1$  \\
$g g \to g\bar{q}q$ & 0 & 0 & $\frac{1}{2}$ & 0 & -1 & 0 & $1$   \\
\vspace{-0.4cm} \\
\hline
\end{tabular}
\end{center}
\caption{Singular spin-summed global Laurent coefficients for the general case.  $\alpha = 1$ is \Ar partitioning and $\alpha = 0$ is GGG partitioning.\label{tab:summed_antennae}}
\end{table}

\subsection{\label{app:globalhel}Construction of Global Helicity-Dependent Antennae}

In this appendix, we will provide details for the construction of the helicity-dependent global antennae.
We will assume that the only 
possible non-zero $gg\to ggg$ global antennae are those which also have corresponding non-zero sector 
antennae.  That is, antennae such as $++ \to ---$ will be set to zero\footnote{These antennae must vanish 
by imposing the collinear limit constraints along with the positivity constraints.}.  Otherwise we will assume 
the antennae are non-zero.  Without loss of generality, all possible non-zero antennae
can be expressed as:
\begin{align*}
++\to  +++ &=& \frac{1}{y_{ij}y_{jk}} + \frac{f(y_{jk})}{y_{ij}}+ \frac{f(y_{ij})}{y_{jk}} \ ,\quad 
++\to  +-+ &=& \frac{1}{y_{ij}y_{jk}} + \frac{g(y_{jk})}{y_{ij}}+ \frac{g(y_{ij})}{y_{jk}}  \\
++\to  -++ &=& \frac{h_1(y_{jk})}{y_{ij}}+ \frac{h_2(y_{ij})}{y_{jk}} \ ,\quad 
++\to  ++- &=& \frac{h_2(y_{jk})}{y_{ij}}+ \frac{h_1(y_{ij})}{y_{jk}} \\ 
+-\to  ++- &=& \frac{1}{y_{ij}y_{jk}} + \frac{a_1(y_{jk})}{y_{ij}}+ \frac{a_2(y_{ij})}{y_{jk}}\ , \quad 
+-\to  +-- &=& \frac{1}{y_{ij}y_{jk}} + \frac{b_1(y_{jk})}{y_{ij}}+ \frac{b_2(y_{ij})}{y_{jk}} \\
+-\to  +-+ &=& \frac{c_1(y_{jk})}{y_{ij}}+ \frac{c_2(y_{ij})}{y_{jk}}\ , \quad 
+-\to  -+- &=& \frac{c_2(y_{jk})}{y_{ij}}+ \frac{c_1(y_{ij})}{y_{jk}}\\  
\end{align*}
for some quadratic polynomials $f,g,h_1,h_2,a_1,a_2,b_1,b_2,c_1,c_2$.  All other antennae
are related by $C$ or $P$ symmetry of QCD.
  Note that an antenna only
has a non-zero soft limit if the helicity of the outer gluons in the antenna are conserved.  

As in the unpolarized case in the previous appendix,
 we consider the configuration of four ordered gluons $i,j,k,l$ and study the limit 
$j \parallel k$.  There are 16 distinct $3\to 4$ gluon splittings which could produce these gluons 
which are not related by $C$ or $P$ which can be used to constrain the form of the functions
defined in the antennae.  
Demanding that the eight splitting functions above reproduce the correct soft and collinear 
limits for each of these $3\to 4$ splittings leads to the requirements that:
\begin{align}\label{eq:glob_collinear}
f(z) &= -f(1-z) \nonumber \\
a_1(z)&=b_2(z)=f(z)\\
a_2(z)&=b_1(z)=g(z)\nonumber\\
h_1(z)&=c_2(z)=\frac{z^3-1}{1-z}-g(1-z) \nonumber
\end{align}
and that the functions $h_2(z)$ and $c_1(z)$ are unconstrained.  
One example of the 
constraints is given by, say, $+++\to ++-+$ splitting.  There are two $2\to3$ splittings that 
contribute to the $j\parallel k$ limit:
\begin{eqnarray}
(++)+&\to& (++-)+ \ ,\nonumber \\
+(++) &\to& +(+-+) \ . \nonumber
\end{eqnarray}
Demanding that these two splittings give the correct collinear limit for $j \parallel k$ 
corresponding to a $+\to +-$ gluon splitting demands that
\begin{equation}
h_1(z)+\frac{1}{1-z}+g(1-z)=\frac{z^3}{1-z} \ .
\end{equation}

In addition to the collinear limit constraints we must also demand that the spin-dependent antennae
sum to reproduce the spin-summed antennae as listed in \tabRef{tab:summed_antennae} as
well as the positivity requirements.  First, considering the spin-summed requirement, the numerators
of the splitting functions must sum appropriately:
\begin{align}\label{eq:spin_summed}
-2-\alpha + (1+2\alpha)z - z^2 &=  f(z)+h_1(z)+a_1(z)+c_1(z) \nonumber\\
&= f(z)+h_2(z)+a_2(z)+c_2(z)\\
&=g(z)+h_2(z)+b_1(z)+c_2(z) \nonumber\\
&=g(z)+h_1(z)+b_2(z)+c_1(z)\ . \nonumber
\end{align}
$\alpha$ is the parameter of the spin-summed global antennae as defined in the previous
appendix.  The positivity requirements can be applied to each antenna function individually, and, in
general, the two collinear limits can be used to constrain each antenna; namely 
$y_{ij}\to 0$, $y_{jk}>0$ and $y_{jk}\to 0$, $y_{ij}>0$.  This leads to the following inequalities:
\begin{align}\label{eq:positivity}
\frac{1}{z}+f(z)&\geq0, \qquad \frac{1}{z}+a_1(z)\geq0, \qquad \frac{1}{z}+a_2(z)\geq0,  \qquad c_1(z)\geq0, \qquad c_2(z)\geq0 \nonumber \\
\frac{1}{z}+g(z)&\geq 0, \qquad \frac{1}{z}+b_1(z)\geq0, \qquad \frac{1}{z}+b_2(z)\geq0,  \qquad h_1(z)\geq0, \qquad h_2(z)\geq0  
\end{align}
where $0<z\leq 1$ in the final-state shower phase space.

Imposing all constraints from \eqRef{eq:glob_collinear}, \eqRef{eq:spin_summed} and 
\eqRef{eq:positivity}, the singular terms in the helicity-dependent global antennae for gluon emission can be written as:
\begin{eqnarray*}
++ \to +++ &=& \frac{1}{y_{ij}y_{jk}}+\frac{(\alpha_1-\alpha+1)-2(\alpha_1-\alpha+1)y_{jk}}{y_{ij}}+\frac{(\alpha_1-\alpha+1)-2(\alpha_1-\alpha+1)y_{ij}}{y_{jk}} \\
++ \to +-+ &=&  \frac{1}{y_{ij}y_{jk}}+ \frac{-(\alpha_1+3)+(2\alpha_1+3-\beta_1)y_{jk}-(\alpha_1+1-\beta_1)y_{jk}^2}{y_{ij}} \\
&& \ + \ \frac{-(\alpha_1+3)+(2\alpha_1+3-\beta_1)y_{ij}-(\alpha_1+1-\beta_1)y_{ij}^2}{y_{jk}} \\
++\to -++ &=& \frac{\beta_1y_{jk}+(\alpha_1-\beta_1)y_{jk}^2}{y_{ij}}\\
++\to ++- &=& \frac{\beta_1y_{ij}+(\alpha_1-\beta_1)y_{ij}^2}{y_{jk}} \\
+-\to ++- &=& \frac{1}{y_{ij}y_{jk}}+\frac{(\alpha_1-\alpha+1)-2(\alpha_1-\alpha+1)y_{jk}}{y_{ij}}\\
&& \ + \ \frac{-(\alpha_1+3)+(2\alpha_1+3-\beta_1)y_{ij}-(\alpha_1+1-\beta_1)y_{ij}^2}{y_{jk}} \\
+-\to +-- &=& \frac{1}{y_{ij}y_{jk}} +  \frac{-(\alpha_1+3)+(2\alpha_1+3-\beta_1)y_{jk}-(\alpha_1+1-\beta_1)y_{jk}^2}{y_{ij}} \\
&& \ + \ \frac{(\alpha_1-\alpha+1)-2(\alpha_1-\alpha+1)y_{ij}}{y_{jk}}\\
+-\to +-+ &=& \frac{\beta_1y_{ij}+(\alpha_1-\beta_1)y_{ij}^2}{y_{jk}}\\
+-\to -+- &=& \frac{\beta_1y_{jk}+(\alpha_1-\beta_1)y_{jk}^2}{y_{ij}}
\end{eqnarray*}
for the parameters $\alpha$, $\alpha_1$ and $\beta_1$.  $\alpha$ is the spin-summed parameter
which can be set appropriately to compare the spin-dependent antennae to the corresponding
spin-summed or unpolarized antennae.  
The constraints on positivity are that $\alpha \geq 0$, $0 \leq \alpha_1 \leq \alpha$ and $\beta_1\geq 0$.
 
We also must impose the constraint that all antennae are positive in the non-singular regions
of phase space.  This requires that finite terms are added to some spin-dependent antennae.
The procedure for determining the non-singular terms will be discussed in~\appRef{app:sector_ant} and 
similar non-singular terms are found in the global case as in the sector case.  These non-singular terms
are included in~\tabRef{tab:glob_antenna_coefficients}.

\subsection{Construction of Helicity-Dependent Sector Antennae \label{app:sector_ant}}
 
In this appendix, we will provide an example of how the sector antennae are constructed from 
their singular limits.  Consider the splitting $g_+ g_+ \to g_+ g_- g_+$ to three final-state gluons $i$, $j$ and $k$, respectively.
The singular terms of the sector antenna for this splitting can be written in the generic form as
\begin{equation}
a(g_+ g_+ \to g_+ g_- g_+) = \frac{1}{y_{ij} y_{jk}} + \frac{f(y_{jk})}{y_{ij}} + \frac{f(y_{ij})}{y_{jk}} \ ,
\end{equation}
for some function $f(z)$.  The form of this antenna is constrained by its soft and collinear limits.  The eikonal
term, $1/y_{ij}y_{jk}$, is required by the existence of a soft limit for the emission of the negative helicity gluon for this splitting.
The other terms are constrained by the form of the collinear limits.  Note that the splitting is symmetric under the interchange
of gluons $i$ and $k$ which demands that the numerator of the terms proportional to $1/y_{ij}$ and $1/y_{jk}$ be 
identical.  Thus, to determine the full form of the singular terms, we only need to consider a single collinear limit.  

In the limit that $i\parallel j$, the antenna must reproduce the splitting function:
\begin{equation}
a(g_+ g_+ \to g_+ g_- g_+) \xrightarrow{i\parallel j} \frac{1}{y_{ij}}P_{g_- g_+\leftarrow g_+}(z)=\frac{1}{y_{ij}}\frac{(1-z)^3}{z} \ ,
\end{equation}
where $z$ is the energy fraction of the emitted negative helicity gluon.  In this limit, $y_{ij}\to 0$ and $y_{jk}\to z$ and this
constrains the function $f(z)$:
\begin{equation}
a(g_+ g_+ \to g_+ g_- g_+) \xrightarrow{i\parallel j} \frac{1}{y_{ij}}\left( \frac{1}{z} + f(z)\right)=\frac{1}{y_{ij}}\frac{(1-z)^3}{z} \ .
\end{equation}
It follows that $f(z) = -3 + 3z -z^2$ and thus the antenna is
\begin{eqnarray}
a(g_+ g_+ \to g_+ g_- g_+) &=& \frac{1}{y_{ij} y_{jk}} - \frac{3}{y_{ij}} -\frac{3}{y_{jk}} + 3 \frac{y_{jk}}{y_{ij}} + 3\frac{y_{ij}}{y_{jk}}
-\frac{y_{jk}^2}{y_{ij}}-\frac{y_{ij}^2}{y_{jk}} \nonumber \\
&+&  \text{ non-singular terms}
\end{eqnarray}

This form of the antenna produces the correct limiting behavior.  However, to be able to use the
antenna in a Markov chain Monte Carlo, it must also have the interpretation as a probability density
and so must be non-negative on all of phase space.  Currently, \Vc only showers the final state,
so we will only consider the final-state phase space for this antenna.  Note that, for example, at the point
$y_{ij} = y_{jk} = 1/2$, the singular terms of the antenna sum to the value $-3$.  Therefore, we must carefully add
non-singular terms to the antenna to guarantee positivity.

To do this, we will find the minima of the antenna on final-state phase space and add the minimal non-singular
terms necessary to guarantee positivity.  By the symmetry of the antenna, the minima lies on the line $y_{ij} = y_{jk}=x$.
Along this line, the derivative of the singular terms of the antenna is
\begin{equation}
\frac{d}{dx}a(g_+ g_+ \to g_+ g_- g_+)_{\text{sing}} = -\frac{2}{x^3}+\frac{6}{x^2}-2 \ .
\end{equation}
Demanding that the derivative be zero at the minima produces a cubic equation with an irrational solution on phase space.
To simplify this, we add the non-singular term $y_{ij}+y_{jk}$ to the original antenna.  This term removes the $-2$ from the derivative
producing a very simple equation to find the minima of the modified antenna.  The minima is located at $x = 1/3$ where the
modified antenna takes the value $-3$.  Therefore, the following antenna is used for the splitting $g_+ g_+ \to g_+ g_- g_+$
in the sector shower in \Vc:
\begin{equation}
a(g_+ g_+ \to g_+ g_- g_+) = \frac{1}{y_{ij} y_{jk}} - \frac{3}{y_{ij}} -\frac{3}{y_{jk}} + 3 \frac{y_{jk}}{y_{ij}} + 3\frac{y_{ij}}{y_{jk}}
-\frac{y_{jk}^2}{y_{ij}}-\frac{y_{ij}^2}{y_{jk}} + 3 + y_{ij} + y_{jk} \ .
\end{equation}
This is non-negative on all of final-state phase space.

\section{Jeppsson 5 Tune Parameters \label{app:j5}}
Note: the Jeppsson 5 parameter set is optimized for use with 
\Vc and to some extent depends on the behavior of that shower model
near the hadronization cutoff. It is therefore not advised to use 
this parameter set directly for standalone \Py 8.

{\small
\begin{verbatim}
! * alphaS
Vincia:alphaSvalue        = 0.139  ! alphaS(mZ) value
Vincia:alphaSkMu          = 1.0    ! Renormalization-scale prefactor
Vincia:alphaSorder        = 1      ! Running order
Vincia:alphaSmode         = 1      ! muR = pT:emit and Q:split
Vincia:alphaScmw          = off    ! CMW rescaling of Lambda on/off

! * Shower evolution and IR cutoff
Vincia:evolutionType      = 1      ! pT-evolution
Vincia:orderingMode       = 2      ! Smooth ordering
Vincia:pTnormalization    = 4.     ! QT = 2pT
Vincia:cutoffType         = 1      ! Cutoff taken in pT
Vincia:cutoffScale        = 0.6    ! Cutoff value (in GeV)

! * Longitudinal string fragmentation parameters
StringZ:aLund             = 0.38   ! Lund FF a (hard fragmentation supp)
StringZ:bLund             = 0.90   ! Lund FF b (soft fragmentation supp)
StringZ:aExtraDiquark     = 1.0    ! Extra a to suppress hard baryons

! * pT in string breakups
StringPT:sigma            = 0.275  ! Soft pT in string breaks (in GeV)
StringPT:enhancedFraction = 0.01   ! Fraction of breakups with enhanced pT
StringPT:enhancedWidth    = 2.0    ! Enhancement factor

! * String breakup flavor parameters
StringFlav:probStoUD     = 0.215   ! Strangeness-to-UD ratio
StringFlav:mesonUDvector = 0.45    ! Light-flavor vector suppression
StringFlav:mesonSvector  = 0.65    ! Strange vector suppression
StringFlav:mesonCvector  = 0.80    ! Charm vector suppression
StringFlav:probQQtoQ     = 0.083   ! Diquark rate (for baryon production)
StringFlav:probSQtoQQ    = 1.00    ! Optional Strange diquark suppression
StringFlav:probQQ1toQQ0  = 0.031   ! Vector diquark suppression
StringFlav:etaSup        = 0.68    ! Eta suppression
StringFlav:etaPrimeSup   = 0.11    ! Eta' suppression
StringFlav:decupletSup   = 1.0     ! Optional Spin-3/2 Baryon Suppression

\end{verbatim}
}

\bibliography{helicity}
\end{document}